\documentclass[journal=jctcce,manuscript=article]{achemso}

\usepackage[version=3]{mhchem} 
\usepackage{upgreek}
\usepackage{xspace}
\usepackage{balance} 
\usepackage{rotating}
\usepackage{url}
\usepackage{multirow}
\usepackage{amsmath}
\usepackage[T1]{fontenc}
\usepackage{color, colortbl}
\usepackage{bm}
\usepackage[table]{xcolor}
\usepackage{dcolumn}
\usepackage{longtable}
\usepackage{nicefrac}
\usepackage{tablefootnote}
\usepackage{threeparttable}
\usepackage{float}
\usepackage{makecell}
\AtBeginDocument{}
\usepackage{chemmacros}
\newcommand{\Lower}[1]{\smash{\lower 1.5ex \hbox{#1}}}

\newcommand{\mc}{\multicolumn}

\usepackage{listings}
\usepackage{physics}

\usepackage{appendix}

\definecolor{mygreen}{rgb}{0,0.6,0}
\definecolor{mygray}{rgb}{0.5,0.5,0.5}
\definecolor{mymauve}{rgb}{0.58,0,0.82}
\definecolor{python_bg}{RGB}{247, 247, 247}
\definecolor{halfgray}{gray}{0.55}
\definecolor{python_frame}{RGB}{207, 207, 207}

\newcommand{\pybest}{\textsc{PyBEST}\xspace}
\newcommand{\Cpp}{C\nolinebreak\hspace{-.05em}\raisebox{.4ex}{\tiny\bf +}\nolinebreak\hspace{-.10em}\raisebox{.4ex}{\tiny\bf +}\xspace}
\newcommand{\Cppeleven}{C\nolinebreak\hspace{-.05em}\raisebox{.4ex}{\tiny\bf +}\nolinebreak\hspace{-.10em}\raisebox{.4ex}{\tiny\bf +}11\xspace}

\newcommand{\ab}{\bar{a}}                                                       
\newcommand{\ib}{\bar{i}}
\newcolumntype{d}[1]{D{.}{.}{#1}}

\author{Katharina Boguslawski}
\email{k.boguslawski@fizyka.umk.pl}
\author{Aleksandra Leszczyk}
\author{Artur Nowak}
\author{Filip Brz\k{e}k}
\author{Piotr Szymon \.Zuchowski}
\affiliation[Institute of Physics, Nicolaus Copernicus University in Toru\'n]
{Institute of Physics, Faculty of Physics, Astronomy and Informatics, Nicolaus Copernicus University in Toru\'n, Grudziadzka 5, 87-100 Torun, Poland}
\author{Dariusz K\k{e}dziera}
\affiliation[Faculty of Chemistry, Nicolaus Copernicus University in Toru\'n]
{Faculty of Chemistry, Nicolaus Copernicus University in Toru\'n, Gagarina 7, 87-100 Torun, Poland}
\author{Pawe{\l} Tecmer}
\affiliation[Institute of Physics, Nicolaus Copernicus University in Toru\'n]
{Institute of Physics, Faculty of Physics, Astronomy and Informatics, Nicolaus Copernicus University in Toru\'n, Grudziadzka 5, 87-100 Torun, Poland}
\email{ptecmer@fizyka.umk.pl}

\title[Pythonic Black-box Electronic Structure Tool (PyBEST). An open-source Python platform for electronic structure calculations at the interface between chemistry and physics.]
 {Pythonic Black-box Electronic Structure Tool (PyBEST). An open-source Python platform for electronic structure calculations at the interface between chemistry and physics}
 
\abbreviations{PyBest, AP1roG, PTa, PTb, LCCD, LCCSD, pCCD, EOM-pCCD, EOM-pCCD+S EOM-pCCD-LCCSD, EOM-pCCD-LCCD, SAPT, quantum chemistry software, Python3}
\begin{document}
 
\begin{abstract}
Pythonic Black-box Electronic Structure Tool (\pybest) represents a fully-fledged modern electronic structure software package developed at Nicolaus Copernicus University in Toruń.
The package provides an efficient and reliable platform for electronic structure calculations at the interface between chemistry and physics using unique electronic structure methods, analysis tools, and visualization. Examples are the (orbital-optimized) pCCD-based models for ground- and excited-states electronic structure calculations as well as the quantum entanglement analysis framework based on the single-orbital entropy and orbital-pair mutual information.  
\pybest is written primarily in the \texttt{Python3} programming language with additional parts written in \Cpp, which are interfaced using \texttt{Pybind11}, a lightweight header-only library. 
By construction, \pybest is easy to use, to code, and to interface with other software packages. 
Moreover, its modularity allows us to conveniently host additional \texttt{Python} packages and software libraries in future releases to enhance its performance. 
The electronic structure methods available in \pybest are tested for the half-filled 1-D model Hamiltonian.  
The capability of \pybest to perform large-scale electronic structure calculations is demonstrated for the model vitamin B$_{12}$ compound. 
The investigated molecule is composed of 190 electrons and 777 orbitals for which an orbital optimization within pCCD and an orbital entanglement and correlation analysis are performed for the first time. 
\end{abstract}

\section{Introduction}
The Pythonic Black-box Electronic Structure Tool (PyBEST) is a modern open-source electronic structure software package developed at Nicolaus Copernicus University (NCU) in Toruń. 
\pybest is written primarily in the \texttt{Python3} programming language (about 90\% of the code) with additional parts written in \Cpp (about 10\% of the code, at least \Cppeleven standard), conveniently interfaced using \texttt{Pybind11}, a lightweight header-only library.~\cite{pybind11} 
The source-code is under the version control (\textsc{git}) and hosted on a \textsc{gitlab} repository server. 
The \pybest project started in 2015 (initially under the name of~\textsc{PIERNIK}) as a spin-off of the \textsc{Horton.v2.0.0} software package developed by T. Verstraelen and coworkers~\cite{horton2.0.0}, to which some of us contributed (K.B. and P.T). 
The newest \textsc{Horton3} package is split into several modules and covers mainly unique features related to Density Functional Approximations (DFAs), integration algorithms, and molecular dynamics.~\cite{horton3.0.0} 
The \pybest development team focuses primarily on unconventional wavefunction models based on the pair Coupled-Cluster Doubles (pCCD) ansatz~\cite{limacher_2013,pccd-orbital-optimization,tamar-pcc}, also known as the Antisymmetric Product of 1-reference orbital Geminal (AP1roG) ansatz in the geminal community, and its extensions.
Thus, the main driving force for the development of \pybest is to model large and complex electronic structures that feature both static/non-dynamic and dynamic electron correlation effects using efficient and reliable wavefunction-based methods. 
On top of that, \pybest also comprises the standard Coupled-Cluster Singles and Doubles (CCSD) approach and its linearized variants (LCCSD and LCCD), the M\"oller--Plesset Perturbation Theory of the second-order (MP2)~\cite{bartlett_2007} and its spin-component scaled variants (SCS-MP2)~\cite{scs-mp2}, as well as the Symmetry Adapted Perturbation Theory of zeroth-order (SAPT0).~\cite{rybak1991}
Over the past few years, \pybest transformed into a new standalone electronic structure software package that is conveniently applicable at the interface between quantum chemistry and physics. 
We should stress, however, that the current version of \pybest uses the original, albeit slightly modified and adjusted \textsc{Horton v.2.0.0} implementation of SCF acceleration techniques such as the commutator-based direct inversion of the iterative subspace (CDIIS) algorithm~\cite{cdiis}, the energy-direct inversion of the iterative subspace (EDIIS) algorithm, the EDIIS2 algorithm (a combination of CDIIS and EDIIS).~\cite{ediis2}
Since we are exploiting a modified version of the \textsc{Horton v.2.0.0} SCF code, all complementary modules required to execute an SCF calculation have been adopted as well.
These include, among others, the occupation model, orbital instances, and the basic structure of the linear algebra factory and the I/O container.
However, neither the SCF implementation of \pybest nor any other module warrants any backward compatibility with \textsc{Horton v.2.0.0}.

\pybest's unique features include the (variational) orbital-optimized pCCD model~\cite{pccd-orbital-optimization,pccd-jctc-2014}, which allows us to optimize molecular orbitals similar as in the Complete Active Space Self-Consistent-Field~\cite{roos_casscf} (CASSCF) procedure but with no restriction on the number of active orbitals or electrons, that is, without the need to define active orbitals spaces. 
Furthermore, pCCD also provides a cost-effective way to obtain an orbital entanglement and correlation analysis form its response density matrices.~\cite{ijqc-entanglement,ijqc-eratum,pawel_pccp2015, pCCD-prb-2016} 
It opens the way to large-scale modeling and a quantum entanglement analysis of complex electronic structures, which feature a significant amount of strong electron correlation effects.~\cite{ps2-ap1rog, pccd-orbital-optimization, pccd-jctc-2014, pawel_pccp2015, pCCD-prb-2016} 
The missing part of the dynamic correlation energy in pCCD can be accounted for using either one of the perturbation theory models~\cite{pccd-PTX, piotrus_pt2,filip-jctc-2019} or a (Linearized) Coupled-Cluster ((L)CC) correction.~\cite{kasia-lcc} 
Both the pCCD and pCCD-LCC ans\"atze have been extended to model excited states using the Equation of Motion (EOM) formalism.~\cite{eom-pccd,eom-pccd-erratum, eom-pccd-lccsd, pawel-yb2, Nowak2019} 
The first version of the code, \pybest v1.0.0, was released on July 1, 2020, and is available free of charge.~\cite{pybestv1.0.0} 
Most recent changes are available at the \pybest project homepage.~\cite{pybest_web}
 
\section{Program overview}\label{sec:overview}
The \pybest software package comprises the Self-Consistent Field (SCF) module for the Hartree--Fock method, the pCCD module including two different orbital optimization protocols, the Perturbation Theory (PT) module, the CC module, the EOM-CC module, and the SAPT module. 
These are augmented with post-processing capabilities such as the Pipek--Mezey 
localization~\cite{pipek-mezey}, an orbital entanglement and orbital-pair correlation analysis~\cite{ijqc-entanglement, ijqc-eratum}, as well as the calculation of the electric dipole and quadrupole moments, where a wrapper is provided for the former.~\cite{helgaker_book} 
\subsection{Program capabilities}\label{sec:capabilities}
The \pybest software package allows us to perform calculations for model Hamiltonians as well as the (non-relativistic) molecular Hamiltonian. 
Furthermore, it is possible to read in the matrix representation of a Hamiltonian generated by an external software package and perform calculations with \pybest (\textit{e.g.}, Hartree--Fock, post-Hartree--Fock, and post-processing).
Alternatively, the matrix representation of a Hamiltonian (expressed in some molecular orbital basis) can be exported in \pybest to be imported by external software suits. 
The communication between different packages is possible through the so-called \path{FCIDUMP} ASCII standard present in, for instance, the \textsc{Molpro}~\cite{molpro2012,molpro-wires}, \textsc{Dalton}~\cite{dalton-2014}, and \textsc{Budapest DMRG}~\cite{dmrg_ors} software packages. 
\subsubsection{The general program structure }\label{sec:general-structure}
The \pybest software package has a well-defined structure that is similar to all electronic structure methods. 
Methods implemented in \pybest operate on and return custom multidimensional array classes that store the actual arrays as their private attributes. Prior to all electronic structure calculations, the user can specify how tensors are represented in \pybest (see also section~\ref{sec:linalg}). These are, then, combined with a specific Hamiltonian and electronic structure method. The user has full control of the initial conditions and computational workflow by specifying all quantities, e.g., basis set, selected one- and two-electron integrals, occupation model, and initial orbitals (see also sections \ref{sec:hamiltonians} and \ref{sec:orbitals} for more details). 
Thus, prior to electronic structure calculations with \pybest, all quantities, like orbitals, occupation models, integrals, etc., need to be defined and initialized. 
\subsubsection{Hamiltonians and basis sets}\label{sec:hamiltonians}
In \pybest v1.0.0, there are two possible choices of Hamiltonians: (i) model Hamiltonians and (ii) the non-relativistic quantum chemical Hamiltonian. 
Currently as model Hamiltonians, \pybest supports the 1-dimensional Hubbard Hamiltonian~\cite{hubbard_model} with and without periodic boundary conditions featuring an adjustable hopping and on-site interaction term.
The non-relativistic quantum chemical Hamiltonian is constructed from a given molecular geometry and (atom-centered) Gaussian basis sets. 
The molecular geometry can be either introduced directly using a \texttt{Python}-based input style or read from a \path{.xyz} file. 
The basis set information can be loaded from \pybest's basis set library or a user-specified file.
Since \pybest interfaces the basis set reader shipped with \texttt{Libint} (using the modern \Cpp API), the basis set has to be provided in the \path{.g94} format.
\begin{lstlisting}[language=Python]
# create a Gaussian basis set for molecular coordinates provided in mol.xyz
gobasis = get_gobasis("cc-pvdz", "mol.xyz")
\end{lstlisting}
\pybest also allows us to add ghost atoms and mid-bond functions (see section~\ref{sec:fragment-module}). 
The complete information about the molecular geometry and basis set (including the molecular orbitals) can be provided directly in \pybest or read from a file that has been either dumped in \pybest's internal \path{.h5} format or generated by some external software in the \path{.molden} and \path{.mkl} formats.
In \pybest, the molecular Hamiltonian is constructed term-wise, that is, the kinetic energy, the nucleus-electron attraction, and the electron repulsion integrals as well as the nuclear repulsion term.
Besides, the overlap integrals of the atom-centered basis set have to be calculated.
\begin{lstlisting}[language=Python]
# compute integrals in the atom-centered Gaussian basis set
kin = compute_kinetic(gobasis)
na = compute_nuclear(gobasis)
nuc = compute_nuclear_repulsion(gobasis)
# overlap matrix of the atom-centered Gaussian basis set
olp = compute_overlap(gobasis)
# dense electron repulsion integrals
er = compute_eri(gobasis)
\end{lstlisting}
The developer version of \pybest also supports the (scalar) relativistic Hamiltonians, like the Douglas--Kroll--Hess Hamiltonian of second order.~\cite{dkh1,dkh2,darek-dirac-limit,reiher_book,reiher2012relativistic,pawel-relativistic-book-chapter-2016}
Instead of the dense representation of the electron repulsion integrals (ERI), the user can also choose Cholesky-decomposed ERI.~\cite{cholesky-review-2011}
The truncation threshold can be changed using the \texttt{threshold} argument (the default value is set to $10^{-3}$).
\begin{lstlisting}[language=Python]
# Cholesky-decomposed electron repulsion integrals
er = compute_cholesky_eri(gobasis, threshold=1e-8)
\end{lstlisting}
\subsubsection{Tensor representations}\label{sec:linalg}
In version v1.0.0, two tensor representations are available: (i) the dense representation, where all elements (including zeros) are stored, and (ii) Cholesky decomposition, where only the electron repulsion integrals are decomposed, while all smaller dimensional objects are stored in their dense representation.
By creating an instance of the selected \texttt{LinalgFactory} class and passing it as an argument during the initialization of quantum chemical methods, the user specifies the preferred tensor representation.
\begin{lstlisting}[language=Python]
# create an instance of the dense linalg factory
lf = DenseLinalgFactory(gobasis.nbasis)
# create an instance of the Cholesky linalg factory
lf = CholeskyLinalgFactory(gobasis.nbasis)
\end{lstlisting}
The linear algebra module is described in more detail in section~\ref{sec:tce}.
\subsubsection{Molecular orbitals and orbital occupation models}\label{sec:orbitals}
To perform a Hartree--Fock or post-Hartree--Fock calculation, a set of (molecular) orbitals, including their occupation model, has to be defined.
In \pybest, these orbitals contain information on the AO/MO coefficient matrix, the orbital occupation numbers, and the orbital energies.
\begin{lstlisting}[language=Python]
# create an empty set of orbitals using the LinalgFactory
orb_a = lf.create_orbital()
\end{lstlisting}
In \pybest, the molecular charge is specified by the number of (singly or doubly) occupied orbitals, that is, the difference between the number of electrons in (singly or doubly occupied) orbitals and the sum of the atomic numbers of all atoms in the molecule.
The current version features three different occupation models.
In most electronic structure calculations, however, only the \texttt{AufbauOccModel} is supported, where the molecular orbitals are filled with respect to the Aufbau principle.
\begin{lstlisting}[language=Python]
# restricted case (six alpha and six beta electrons)
occ_model = AufbauOccModel(6)
# unrestricted case (four alpha and three beta electrons)
occ_model = AufbauOccModel(4, 3)
\end{lstlisting}
The fixed occupation model and the so-called Fermi occupation model~\cite{fermi-occ-model-99} represent alternative choices.  
\subsubsection{The Hartree--Fock module}\label{sec:hf}
To perform an SCF calculation for the Hartree--Fock method, all quantities mentioned above have to be defined, that is, some (molecular) Hamiltonian including the molecular geometry and basis set, a \texttt{LinalgFactory} instance, a set of (molecular) orbitals, and an occupation model. 
Both the restricted and unrestricted variants of the Hartree--Fock method are implemented. 
\pybest offers a convenient wrapper to perform (restricted or unrestricted) Hartree--Fock (RHF and UHF) calculations in an automatic manner. 
This wrapper combines all terms of the Hamiltonian, generates some initial guess orbitals, chooses a default DIIS solver, and a default convergence threshold.
Upon convergence, all Hartree--Fock output data required for restarts and post-processing is returned as an instance of a \pybest-specific container class.
\begin{lstlisting}[language=Python]
# create an instance of the RHF class
hf = RHF(lf, occ_model)
# start an RHF calculation
hf_ = hf(kin, na, er, nuc, olp, orb_a)
\end{lstlisting}
A common feature of \pybest is that the order of the arguments in the function call does not matter as \pybest exploits internally defined labels for all tensors.
Any RHF (or UHF) calculations can be restarted from an internal checkpoint file featuring the \path{.h5} extension.
This can be achieved by using the \texttt{restart} keyword and providing the path to the checkpoint file in the function call. By default, all internal checkpoint files are stored in the \texttt{pybest-results} directory using method-specific naming conventions.
\begin{lstlisting}[language=Python]
# restart an RHF calculation
hf_ = hf(kin, na, er, nuc, olp, orb_a, restart="pybest-results/checkpoint_scf.h5")
\end{lstlisting}
In addition to standard restarts from checkpoint files, \pybest allows for restarts from perturbed orbitals by swapping orbital pairs or performing manual orbital rotations (Givens rotations), localized orbitals, as well as random unitary rotations of all molecular orbitals. 
The SCF procedure in \pybest can be performed with or without acceleration techniques. 
Possible choices of DIIS solvers include the CDIIS, EDIIS, and EDIIS2 methods~\cite{cdiis,ediis2} and can be invoked by the \texttt{diis} keyword.
After the SCF algorithm is converged, the final data can be post-processed (\textit{e.g.}, orbital localization) or passed as input to post-HF methods. 
\subsubsection{The general structure of post-HF modules}\label{sec:posthf}
All post-HF modules in \pybest have a similar input structure.
To optimize the wavefunction of any post-HF method, the corresponding module requires a Hamiltonian and some orbitals including the overlap matrix (for restart purposes) as input arguments. 
Note that only the Hamiltonian terms have to be passed explicitly. 
All remaining information is stored in the HF output container \path{hf_}.  
\begin{lstlisting}[language=Python]
# create an instance of a post-Hartree-Fock method (here called PostHF)
posthf = PostHF(lf, occ_model)
# start the calculation through a function call using the results of
# a previous RHF calculation stored in hf_
posthf_output = posthf(kin, na, er, hf_)
\end{lstlisting}
In the above example, \texttt{PostHF} is some post-HF flavor (see below).
Thus, the input structure is similar to an HF calculation, except that only all Hamiltonian terms have to be passed explicitly.
All post-HF modules support only spin-restricted orbitals and both the \path{DenseLinalgFactory} and \path{CholeskyLinalgFactory}. 
Furthermore, \pybest supports frozen core orbitals, that is, a set of occupied orbitals that is excluded in post-HF calculations as they are assumed to be doubly occupied. 
This feature is particularly useful when modeling heavier elements for which the core basis functions are generally not optimized for correlated calculations. 
\subsubsection{The pCCD module}\label{sec:pccd-module}
\pybest supports some unique wavefunction models based on the pCCD ansatz. 
The pCCD model represents an efficient parameterization of the Doubly Occupied Configuration Interaction (DOCI) wavefunction,~\cite{doci} but requires only mean-field computational cost in contrast to the factorial scaling of traditional DOCI implementations.~\cite{limacher_2013,tamar-pcc} 
The pCCD wavefunction ansatz can be rewritten in terms of one-particle functions as a fully general pair-Coupled-Cluster wavefunction,
\begin{equation}\label{eq:pccd}                                               
|{\rm pCCD}\rangle = \exp \left (  \sum_{i=1}^{\rm occ} \sum_{a=1}^{\rm virt} t_i^a a_a^{\dagger}  a_{\bar{a}}^{\dagger}a_{\bar{i}} a_{i}  \right )| 0 \rangle = e^{\hat{T}_{\rm p}} | 0 \rangle,
\end{equation}     
where $a^\dagger_p$ and $a_p$ ($a^\dagger_{\bar{p}}$ and $a_{\bar{p}}$) are the electron creation and annihilation operators for $\alpha$ ($\beta$) electrons, $| 0 \rangle$ is some reference determinant, $\{t_i^a\}$ are the electron-pair amplitudes, and $\hat{T}_{\rm p} = \sum_{i=1}^{\rm occ} \sum_{a=1}^{\rm virt} t_i^a a_a^{\dagger}  a_{\bar{a}}^{\dagger}a_{\bar{i}} a_{i}$ is the electron-pair excitation operator that excites an electron pair from an occupied $(i\ib)$ to a virtual orbital $(a\ab)$ with respect to $| 0 \rangle$.~\cite{limacher_2013}
\pybest supports conventional pCCD calculations with a RHF reference function or the orbital-optimized variant (OO-pCCD).~\cite{pccd-orbital-optimization}
\begin{lstlisting}[language=Python]
# create an instance of RpCCD
pccd = RpCCD(lf, occ_model)
# pCCD calculations
pccd_ = pccd(kin, na, er, hf_)
\end{lstlisting}
All results of a pCCD calculation (\textit{e.g.}, electron-pair amplitudes, electronic energies, etc.) are stored as attributes in the \texttt{pccd$\_$} container.

The pCCD model ensures size-extensivity by construction, requires, however, the optimization of the one-particle basis functions to satisfy size-consistency. 
\pybest features two different orbital optimization protocols: (i) variational orbital optimization~\cite{pccd-orbital-optimization} and (ii) the projected-seniority-two orbital optimization in the commutator formulation (PS2c).~\cite{pccd-jctc-2014}
By default, the variational orbital optimization protocol is selected.
\begin{lstlisting}[language=Python]
# create an instance of ROOpCCD
pccd = ROOpCCD(lf, occ_model)
# (variational) OO-pCCD calculation
pccd_ = pccd(kin, na, er, hf_)
\end{lstlisting}
Within the variational orbital optimization scheme, \pybest performs the calculation of the pCCD response 1- and 2-particle reduced density matrices (1-RDM and 2-RDM), $\gamma^{p}_{q}$ and $\Gamma^{pq}_{rs}$, respectively.
Both RDMs are stored as attributes in the \texttt{pccd$\_$} container. 
The pCCD 1-RDM is diagonal and is calculated from~\cite{ijqc-entanglement,ijqc-eratum}
\begin{equation}\label{eq:1-rdm}                                                                  
    \gamma_p^p = \langle \Psi_0| (1+\hat{\Lambda}) a^\dagger_p a_p | \textrm{pCCD} \rangle,
\end{equation}
where $\hat{\Lambda}$ contains the de-excitation operator, 
\begin{equation}\label{eq:lamba-eq}                
    \hat{\Lambda} = \sum_{ia} \lambda_i^a (a^\dagger_i a^\dagger_{\bar{i}} a_{\bar{a}} a_a - t_i^a).
\end{equation}
The eigenvalues of the response 1-RDM are the pCCD natural occupation numbers.~\cite{pccd-jctc-2014, pawel_pccp2015} 
The response 2-RDM is defined as                                                
\begin{equation}\label{eq:2-rdm}                                                                      
    \Gamma^{pq}_{rs} = \langle \Psi_0| (1+\hat{\Lambda})a^\dagger_p a^\dagger_{q}  a_{s} a_r| \textrm{pCCD} \rangle.
\end{equation} 
Since most of the elements of $\Gamma^{pq}_{rs}$ are zero by construction, only the non-zero elements of the response 2-RDM are calculated in \pybest, which include $\Gamma^{pq}_{pq}=\Gamma^{p\bar{q}}_{p\bar{q}}$ ($\forall \, p \neq q$) and $\Gamma^{p\bar{p}}_{q\bar{q}}$. 
By default, the orbital optimizer exploits a diagonal approximation to the exact orbital Hessian.
The exact orbital Hessian can be calculated separately. 
However, one should keep in mind that the latter is computationally expensive and hence limited to small and moderate system sizes.
To optimize an orbital rotation step, the trust-region or backtracking algorithms can be employed. 
The former is used by default and combined with Powell's double-dogleg approximation of the trust region step.~\cite{double-dogleg} 
This particular setup proved to work best for most of the investigated systems. 
Alternatively, the user can employ the preconditioned conjugate gradient and Powell's single dogleg step optimization algorithms.~\cite{steihaug1983conjugate, powell1970} 
Orbital optimization within pCCD allows us to obtain qualitatively correct potential energy surfaces and captures a large fraction of strong electron correlation effects.~\cite{pawel_jpca_2014, ps2-ap1rog, pccd-orbital-optimization, pCCD-prb-2016, ola-book-chapter-2019}
Currently, the pCCD module is limited to closed-shell systems only.
Various open-shell extensions are currently under development and will be available in future releases of \pybest.
\subsubsection{The CC module}\label{sec:cc-module}
\pybest supports standard CCD and CCSD as well as their linearized variants, that is, LCCD and LCCSD, on top of a (restricted) Hartree--Fock reference wavefunction. The LCCSD method is equivalent to CEPA(0).
All conventional CC calculations can be invoked similarly by creating an instance of the corresponding CC class.
\begin{lstlisting}[language=Python]
# create an instance of RCCSD
ccsd = RCCSD(lf, occ_model)
# CCSD calculation
ccsd_ = ccsd(kin, na, er, hf_)
\end{lstlisting}
Note that all CC models in \pybest also support non-canonical orbitals.
Thus, the \texttt{hf$\_$} input container can be substituted by any (converged) reference determinant.
In addition to the conventional CEPA(0) approximation, the LCC correction can also be combined with a pCCD reference function (with and without orbital optimization), resulting in the pCCD-LCCD and pCCD-LCCSD approaches.~\cite{kasia-lcc} 
To distinguish between the various linearized CC models, \pybest v1.0.0 uses a specific class name convention: \path{RHFLCCD}, \path{RHFLCCSD}, \path{RpCCDLCCD}, and \path{RpCCDLCCSD}, respectively.
\begin{lstlisting}[language=Python]
# create an instance of RpCCDLCCSD
lccd = RpCCDLCCSD(lf, occ_model)
# pCCD-LCCSD calculation with a pCCD reference function
lccd_ = lccd(kin, na, er, pccd_)
\end{lstlisting}
Besides, \pybest allows to perform a standard CC calculations on top of pCCD orbitals. 
All CC calculations in \pybest are restartable from \texttt{.h5} checkpoint files. 
By default, the \texttt{krylov} solver as implemented in \texttt{scipy} is used.~\cite{scipy1.0}
For all LCC flavors, a Perturbation-based Quasi-Newton (\texttt{pbqn}) solver is also provided.
For difficult cases, however, the \texttt{krylov} solver represents a better alternative, despite its slow convergence. 

Furthermore, the $\Lambda$ equations for all LCC models can be solved.
The corresponding $\lambda$ amplitudes are then used to construct the response RDMs of the LCC wavefunction.
In the case of a pCCD reference function, the correlation part can be determined from
\begin{align}\label{eq:responsedms-corr}
(\Gamma^{p\ldots}_{t\ldots})^{\rm corr}=\bra{\Phi_0}(1+\Lambda^\prime) & \{ e^{-\hat{T}^\prime -\hat{T}_p} \{\hat{a}^\dagger_p\ldots a_t \}  \nonumber \\
&e^{\hat{T}_p+\hat{T}^\prime} \}_{L^\prime} \ket{\Phi_0},
\end{align}
where $\hat{T}^\prime$ ($\Lambda^\prime$) contains at most double (de-)excitations, that is $\hat{T}^\prime=\hat{T}_1+\hat{T}_2^\prime$ or $\hat{T}^\prime=\hat{T}_2^\prime$ ($\Lambda^\prime=\Lambda_1+\Lambda_2^\prime$ or $\Lambda^\prime=\Lambda_2^\prime$), respectively, and
\begin{equation}\label{eq:de-excitation}
\Lambda_n^\prime = \frac{1}{(n!)^2} \sum_{ij\ldots}\sum_{ab\ldots}{}^\prime \lambda^{ij\ldots}_{ab\ldots}{i^\dagger a j^\dagger b\ldots}
\end{equation}
is the de-excitation operator, where all electron-pair de-excitation are to be excluded as they do not enter the {LCC} equations (indicated by the "$\prime$").
Since we work with a linearized coupled-cluster correction, all broken-pair excitation appear at most linear in eq.~\eqref{eq:responsedms-corr} (labeled by the subscript $L^\prime$).
The total $N$-RDM is the sum of the reference contribution, the leading correlation contribution eq.~\eqref{eq:responsedms-corr}, and all lower-order correlation contributions,  
\begin{equation}\label{eq:responsedms}
\Gamma^{p\ldots}_{t\ldots}= (\Gamma^{p\ldots}_{t\ldots})^{\rm ref}+(\Gamma^{p\ldots}_{t\ldots})^{\rm corr} + \{(\Gamma^{p\ldots}_{t\ldots})^{\rm corr}_{(N-1,\ldots,1)}\} ,
\end{equation}
where the last term indicates all possible lower-order ($N-1,\ldots,1$) correlation contributions to the $N$-RDM in question.
For more details see also Ref.~\citenum{artur-entanglement-pccd-lccsd}.
The current version of \pybest automatically calculates (selected elements) of the response 1-, 2-, 3-, and 4-particle RDMs that are required for all supported post-processing schemes (see section~\ref{sec:analysis-and-visualization}) when the $\Lambda$ equations are to be solved.
This can be invoked by setting the keyword argument \texttt{l=True}.

\begin{lstlisting}[language=Python]
# creat an instance of pCCD-LCCSD
lccsd = RpCCDLCCSD(lf, occ_model)
# also solve the corresponding Lambda equations
lccsd_ = lccsd(kin, na, er, pccd_, l=True)
\end{lstlisting}
The developer version of \pybest also features tailored coupled cluster methods, like the frozen-pair CCSD flavours.~\cite{frozen-pccd, veis2016} 
\subsubsection{The EOM-CC module}\label{sec:eom-cc-module}
The released version of \pybest allows us to calculate electronically excited states using the EOM-pCCD and EOM-pCCD+S approaches.~\cite{eom-pccd, eom-pccd-erratum} 
\begin{lstlisting}[language=Python]
# create an instance of REOMpCCD
eom = REOMpCCD(lf, occ_model)
# calculate 3 lowest-lying roots
eom_ = eom(kin, na, er, pccd_, nroot=3)
\end{lstlisting}
The EOM-pCCD ansatz includes only the electron-pair excitations in the EOM ansatz while the EOM-pCCD+S flavor comprises both singles and electron-pair excitations. 
The developer version of \pybest also supports the EOM-pCCD-LCCD and EOM-pCCD-LCCSD variants. 
More details about those methods are available in Ref.~\citenum{eom-pccd-lccsd}. 
\subsubsection{The perturbation theory module}\label{sec:pt-module}
\pybest features perturbation theory-based calculations on top of a (canonical) restricted Hartree--Fock and a pCCD reference function.
For the former, \pybest offers the conventional M\o{}ller--Plesset perturbation theory model of second order.
\begin{lstlisting}[language=Python]
# create an instance of the RMP2 class 
mp2 = RMP2(lf, occ_model)
# perform conventional MP2 calculation on RHF canonical orbitals
mp2_ = mp2(kin, na, er, hf_)
\end{lstlisting}
The MP2 module also supports the calculation of (relaxed and unrelaxed) 1-particle reduced density matrices (1-RDM) and the corresponding natural orbitals.
Besides the conventional MP2 implementation, the Spin-Component-Scaled (SCS) variant is also implemented including the corresponding (relaxed and unrelaxed) 1-RDM.
The same-spin and opposite-spin scaling factors are defined using the keyword arguments \path{fss} and \path{fos} in the function call.
Furthermore, the contributions of single excitations can be accounted for through the \path{singles} keyword argument.   
\begin{lstlisting}[language=Python]
# restricted MP2 calculation with single and double excitations
mp2_ = mp2(kin, na, er, hf_, singles=True)
\end{lstlisting}
While single excitations have no effect on the MP2 energy calculated on top of the canonical Hartree--Fock reference, this is no longer the case for non-canonical orbitals like pCCD-optimized orbitals.  

Finally, \pybest provides various unique perturbation theory models with a pCCD reference function that goes beyond the MP2 standard.~\cite{pccd-PTX}
Specifically, these perturbation theory corrections offer different choices for the zeroth-order Hamiltonian, perturbation, dual space, and projection manifold. 
All possible combinations of these degrees of freedom lead to the development of the so-called PT2X models.  
Possible choices for PT2X are: (i) PT2SDd (single determinant dual state and diagonal one-electron zero-order Hamiltonian), (ii) PT2MDd (multi determinant dual state and diagonal one-electron zero-order Hamiltonian), (iii) PT2SDo (single determinant dual state and off-diagonal one-electron zero-order Hamiltonian), (iv) PT2MDo (multi determinant dual state and off-diagonal one-electron zero-order Hamiltonian), and (v) PT2b.~\cite{piotrus_pt2, pccd-PTX}  
By default, the projection manifold is restricted to double excitations but single excitations can be accounted for as well (again using the \path{singles} keyword argument). 
\begin{lstlisting}[language=Python]
# create an instance of the PT2MDo class
pt2 = PT2MDo(lf, occ_model)
# perform PT2MDo calculations including single and double excitations
pt2_ = pt2(kin, na, er, pccd_, singles=True)
\end{lstlisting}
The other perturbation theory corrections can be invoked by creating instances of the perturbation theory classes (i) to (v).
\subsubsection{Fragment-based calculations}\label{sec:fragment-module}
\pybest provides a flexible interface to handle atomic basis sets such that dummy or ghost atoms and active molecular fragments can be easily defined. 
Specifically, all dummy/ghost atoms, as well as active fragments have to be passed to the basis set reader function.
Both ghost/dummy atoms and active fragments can be used as either joint or separate arguments.
To specify the former, the \path{dummy} argument has to be used, where all dummy/ghost atoms are indicated as a \path{list} containing the indices of all atoms in question.
The atoms are indexed with respect to their order in the \path{.xyz} file (\texttt{Python} indexing convention).
\begin{lstlisting}[language=Python]
# set first (0) and second (1) element as dummy/ghost atom
gobasis = get_gobasis("cc-pvdz", "mol.xyz", dummy=[0,1])
\end{lstlisting}
Active fragments are defined in a similar manner.
\begin{lstlisting}[language=Python]
# molecule containing atoms 3, 4, and 5 as active fragment 
gobasis = get_gobasis("cc-pvdz", "mol.xyz", active_fragment=[3,4,5])
\end{lstlisting}
The inactive fragment will be neglected during the construction of the basis set and molecular Hamiltonian.
Such fragment-based calculations are particularly useful in the SAPT module, where the molecular Hamiltonian is partitioned into (presumably) weakly interacting fragments~\cite{jeziorski1994,patkowski-sapt-review-2020}
\begin{equation}
  H = H_{\textrm {A}} + H_{\textrm {B}} + V.  
\end{equation}
The first two terms in the above equation correspond to the Hamiltonians of the monomers A and B, respectively, and $V$ is the interaction between them. 
Such a partitioning defines the perturbation series in powers of $V$
with a $\Psi_{\rm A} \Psi_{\rm B}$ zeroth-order wavefunction.
\pybest's SAPT module currently supports the so-called SAPT0 approximation which neglects the effects of the intramonomer correlation energy on the interaction energy components. In other words, the zeroth-order wavefunction is  a product of Slater determinants.~\cite{rybak1991} 
\begin{lstlisting}[language=Python]
# create an instance of SAPT0
sapt0 = SAPT0(monA, monB)
# SAPT0 calculations
sapt0(monA, monB, dimer)
\end{lstlisting}
\pybest is shipped with a utility function \path{sapt_utils.prepare_cp_hf} that automatically calculates all necessary ingredients for SAPT0 calculations containing the fragments \path{monA} and \path{monB} as well as the supermolecule \path{dimer}.
Furthermore, this wrapper allows us to perform the dimer centered basis set (DCBS) counter-poised corrected Restricted Hartree--Fock calculations in an automatic manner. 
Each correction to the SAPT0 interaction energy has a physical meaning (see Refs.~\citenum{jeziorski1994, patkowski-sapt-review-2020} for more details) and can be accessed term-wise through a dictionary.
\begin{lstlisting} [language=Python]
# SAPT0 dictionary
corrections = sapt0_solver.result
# term-wise contributions to the interaction energy
e10_elst = corrections["E^{(10)}_{elst}"]
e10_exch = corrections["E^{(10)}_{exch}(S^2)"]
e20_ind = corrections["E^{(20)}_{ind},unc"]
e20_exch_ind = corrections["E^{(20)}_{exch-ind}(S^2),unc"]
e20_disp = corrections["E^{(20)}_{disp},unc"]
e20_exch_disp = corrections["E^{(20)}_{exch-disp}(S^2),unc"]
\end{lstlisting}
All implemented expressions are based on the molecular orbital formulation with the $S^2$ approximation for the exchange terms.
\subsubsection{Post-processing: orbital-based analysis and visualization}\label{sec:analysis-and-visualization}
The released version of \pybest supports various post-processing options of electronic wavefunctions and Hamiltonians.
These include dumping orbitals and Hamiltonians to disk using various file formats, (Pipek-Mezey) localization, the calculation of electric dipole moments, and an orbital-based entanglement and correlation analysis.
To visualize molecular orbital, they can be dump to disk within the \path{.molden} format and then visualized using some orbital visualization program.
\pybest interfaces \texttt{libint}'s export molden feature.
In order to dump molecular orbitals to a \path{.molden} file, the \path{to_file} method implemented in the \path{IOData} container can be used.
\begin{lstlisting}[language=Python]
# dump pCCD orbitals stored in the pccd_ container
# update container to contain the atomic basis set
pccd_.gobasis = gobasis
# write molden file
pccd_.to_file("my_pccd_orbitals.molden")
\end{lstlisting}
Note that the \path{pccd_} container has to be updated to include the basis set information explicitly.

Orbital entanglement and correlation represent a unique feature of \pybest and allow us to dissect various wavefunction models using the picture of interacting orbitals.~\cite{rissler2006,barcza2014entanglement,ijqc-entanglement,ijqc-eratum}
This allows us to quantify the interaction between orbitals and to understand chemical processes.
The interaction between orbitals can be determined using concepts of quantum information theory. 
Specifically, to measure the entanglement between one orbital and remaining sets of orbitals, \pybest uses the single-orbital entropy $\rm {s(1)_i}$, defined as follows
\begin{equation}\label{eq:single-entropy_1}
s(1)_i = -\sum_{\alpha=1}^{4} \omega_{\alpha;i}\ln(\omega_{\alpha;i}),
\end{equation}
where $\omega_{\alpha;i}$ are the eigenvalues of the one-orbital reduced density matrix (1-ORDM).
Analogously, the two-orbital entropy $\rm {s_{i,j}}$ is constructed as
\begin{equation}\label{eq:single-entropy_2}
s_{i,j} = -\sum_{\alpha=1}^{16} \omega_{\alpha;i,j}\ln(\omega_{\alpha;i,j}),
\end{equation}
and quantifies the interaction of an orbital pair $i$,$j$, and all other orbitals. The $\omega_{\alpha;i,j}$ are
the eigenvalues of the two-orbital RDM with all possible variants of states for spatial orbitals.
The above quantities are exploited to calculate the orbital-pair mutual information,
\begin{equation}\label{eq:mutal info}
I_{i|j} = s_i + s_j - s_{i,j},
\end{equation}
which determines the correlation between the orbital pair $i$ and $j$ embedded in the environment of all other orbitals.
The matrix elements of the 1- and 2-ORDMs can be expressed in terms of the $N$-particle RDMs (see, for instance, Refs.~\citenum{ijqc-entanglement} and~\citenum{ijqc-eratum}).
The orbital-pair mutual information allows us to dissect electron correlation effects into different contributions and (together with the single-orbital entropy) represents a useful tool to define optimal and balanced active spaces.~\cite{entanglement_letter,ijqc-entanglement,cuo_dmrg,corinne_2015,roland-runo,stein2016,boguslawski-plutonium-oxides-2017,ola-pccp-2019}

The released version of \pybest supports an orbital entanglement and correlation analysis for the pCCD and LCC wavefunction models.
\begin{lstlisting}[language=Python]
# orbital entanglement module
# pCCD
entanglement = OrbitalEntanglementRpCCD(lf, pccd_)
entanglement()
# pCCD-LCCSD
entanglement = OrbitalEntanglementRpCCDLCC(lf, lccsd_)
entanglement()
\end{lstlisting}
The entanglement module dumps all output data to separate files containing the single-orbital entropy, orbital-pair mutual information, and the eigenvalue spectra (including the eigenvectors) of the 1- and 2-ORDMs.
\pybest is shipped with a script \path{pybest-entanglement.py} that allow us to visualize the single-orbital entropy and orbital-pair mutual information in separate graphs (see appendix~\ref{appendix:vitb12} for examples).
\subsection{Implementation details}\label{sec:implementation}
\subsubsection{The \textsc{Libchol} library}
\pybest is shipped with a standalone \Cpp library \path{libchol} that provides the routines to generate Cholesky decomposed electron repulsion integrals (ERIs)~\cite{cholesky-review-2011}.
Specifically, \path{libchol} is partially optimized and constitutes a parallelized part of the \pybest software package.
It, however, represents an optional library, and hence all modules implemented in \pybest run without it.
As \pybest, the \path{libchol} library uses the modern \Cpp API of \texttt{Libint}~\cite{libint} in order to generate blocks of the exact four-center ERIs exploiting MKL's basic linear algebra subprograms \path{BLAS-2} and \path{BLAS-3} for the decomposition algorithm.~\cite{mkl}
While the \path{libchol} library is at an early stage of its development, it provides already a core functionality for the decomposition of ERIs of the orbital basis, 
\begin{equation}\label{eq:choleski-eri}
	( \mu \nu | \rho \sigma ) \approx ( \mu \nu | P ) ( P | \rho \sigma),
\end{equation}
where $P$ is the \textit{ad-hoc} generated Cholesky auxiliary basis set.
The current implementation of \path{libchol} works exclusively in the in-core memory scheme, and thus is limited to medium-size systems with up to, let's say, 1000 basis functions depending on the chosen decomposition threshold. 
Future work includes the out-of-core algorithm and further optimizations allowing us to reduce the required in-core memory and bandwidth.
\subsubsection{The Linear Algebra Factory (LinalgFactory) and the tensor contraction engine}\label{sec:tce}
The \path{linalg} module contains two types of tensor representations: dense tensors and Cholesky-decomposed ERIs, which represent a special type of 4-index objects in \pybest.
All lower-dimensional tensors are always represented as dense \path{numpy} arrays.
\begin{lstlisting}[language=Python]
# dense electron repulsion integrals
dense_er = compute_eri(gobasis)
# cholesky-decomposed electron repulsion integrals
cholesky_er = compute_cholesky_eri(gobasis, threshold=1e-8)
# create two-dimensional array filled with random numbers
operand = DenseTwoIndex(gobasis.nbasis)
operand.randomize()
\end{lstlisting}

Algebraic operations, like addition, tensor contraction, and slicing, are implemented as methods.
Specifically, \pybest's tensor contraction and slicing are abstract layers that provide a standardized interface to various implementations of tensor contraction operations. They are independent of the internal representation of the many-index objects. Thus, the syntax for \texttt{DenseFourIndex} and \texttt{CholeskyFourIndex} classes is the same even if their internal representation is different as a \texttt{CholeskyFourIndex} class stores three-index arrays. 

\begin{lstlisting}[language=Python]
# The following four lines provide the same result (within the accuracy of the Cholesky Decomposition)
A = dense_er.contract("abcd,ab", operand)
B = cholesky_er.contract("abcd,ab", operand)
C = operand.contract("ab,abcd", dense_er)
D = operand.contract("ab,abcd", cholesky_er)
\end{lstlisting}

Since there are many options to perform these kind of algebraic operations (both in \path{Python} and \pybest), the contract method offers a unified interface for \path{opt_einsum.contract},\cite{opt_einsum-2018} \path{numpy.tensordot}, \path{numpy.einsum},\cite{numpy-2020,psi4numpy} and \pybest's BLAS-optimized operations. 
The latter ones are implemented for some bottleneck operations that are complementary to the automatic procedures of the external libraries.

\begin{lstlisting}[language=Python]
# example of application of different tensor contraction engines
# opt_einsum.contract
E = cholesky_er.contract("abcd,cb->ad", operand, select="opt_einsum")
# numpy.tensordot
F = cholesky_er.contract("abcd,cb->ad", operand, select="td") 
# numpy.einsum
G = cholesky_er.contract("abcd,cb->ad", operand, select="einsum") 
# BLAS-optimized operation
H = cholesky_er.contract("abcd,cb->ad", operand, select="blas")
\end{lstlisting}

\pybest is also shipped with different \path{python}-based algorithms to perform a 4-index transformation.
It further provides a convenient wrapper \path{transform_integrals} that automatically transforms all one- and two-electron integrals into the molecular orbital basis.
This transformation can be performed either for restricted or unrestricted orbitals.
Note that the order of the arguments does not matter.

\begin{lstlisting}[language=Python]
# perform an AO/MO transformation of all one- and two-electron integrals
# restricted orbitals
ti = transform_integrals(kin, na, er, orb_a)
# equivalent to
ti = transform_integrals(orb_a, kin, er, na)
# unrestricted orbitals (alpha, beta)
ti = transform_integrals(kin, na, er, orb_a, orb_b)
\end{lstlisting}

By default, \pybest exploits \path{numpy.tensordot} to perform the AO/MO transformation.
Other options ("tensordot" or "einsum") are also possible and are steered using the \path{indextrans} keyword argument.
For both flavors, \pybest successively transforms each index.
For the option "tensordot", the lower-level routine looks as follows, where the transformed integrals are stored in-place,
\begin{lstlisting}[language=Python]
# PyBEST's lower-level routine to perform a 4-index transformation
# due to the way tensordot works, the order of the dot products is
# not according to literature conventions
# for restricted orbitals, we have orb0=orb1=orb2=orb3
self._array[:] = np.tensordot(
    ao_integrals._array, orb0.coeffs, axes=([0], [0])
)
self._array[:] = np.tensordot(
    self._array, orb1.coeffs, axes=([0], [0])
)
self._array[:] = np.tensordot(
    self._array, orb2.coeffs, axes=([0], [0])
)
self._array[:] = np.tensordot(
    self._array, orb3.coeffs, axes=([0], [0])
)
\end{lstlisting}

\subsection{Code documentation and distribution}\label{sec:doc-and-installation}
The documentation of \pybest is generated from sources in the reStructuredText format using \textsc{Sphinx}~\cite{sphinx} and shipped with the code.~\cite{pybestv1.0.0} The most recent documentation is also available on the \pybest homepage.~\cite{pybest_web} 
The documentation covers basic information about the program, license information, how to install it on different operating systems (including dependencies), and a detailed user manual with numerous illustrative examples of how to use the code and its unique features.    
Furthermore, to improve the understanding of larger parts of the code, such as classes, modules, and functions, we use \texttt{Python} \texttt{Docstrings}.
This practice greatly improves code readability and facilitates familiarizing new developers with the code. 
Similar to the source code, the documentation is under version control (\textsc{Git}) and hosted on a \textsc{GitLab} repository server. 

The \pybest software package is distributed in the source code form under an open-source GNUv3 (General Public License) license. 
The source code can be easily downloaded from the \textsc{Zenodo} platform~\cite{zenodo}, where the software package has been minted a permanent DOI number, and its metrics are available. 
A future \pybest release will be assigned to a different DOI number and shipped with a version-specific documentation. 
Additional information is available on the \pybest project homepage.~\cite{pybest_web}
\section{Example calculations using \pybest }\label{sec:results}
\subsection{Electronic structure calculations on the 1-D Hubbard Hamiltonian}\label{sec:hubbard-results}
The 1-D Hubbard Hamiltonian represents a useful model for assessing the accuracy and reliability of newly developed electronic structure methods.~\cite{hubbard-gustavo_2013,pccd-orbital-optimization,tamar-pcc,pCCD-prb-2016} 
Modifying the repulsive on-site interaction parameter ${U}$ allows us to control the strength and nature of electron correlation, where strong correlation dominates for larger $U/t$ values. 
Table~\ref{tbl:hubbard} summarizes the performance of various quantum chemistry models available in \pybest in predicting the total and correlation energies for selected values of $U/t$. 
We focus on the half-filled 1-D Hubbard Hamiltonian with 50 sites and periodic boundary conditions, for which the exact results can be determined by solving the Lieb-Wu equations.~\cite{lieb-wu} 
An example \pybest input file is shown in Appendix~\ref{appendix:hubbard}.
Specifically, we investigate three $U/t$ values, namely $U/t=1$, $U/t=2$, and $U/t=3$.
This mimics the (strong) electron correlation regime usually encountered in molecular systems.
It is evident from Table~\ref{tbl:hubbard} that the OO-pCCD method combined with one of the dynamic energy corrections provides reliable energies, even for cases where the CCSD method fails to converge. 
For a detailed discussion on the performance of various OO-pCCD-LCC approaches for the 1-D Hubbard model, we refer the reader to Ref.~\citenum{artur-entanglement-pccd-lccsd}. 
\begin{table}
\footnotesize{
\begin{center}
   \caption{Performance of various wavefunction models available in \pybest for the half-filled 1-D Hubbard model Hamiltonian for 50 sites with periodic boundary conditions and different values of the repulsive on-site interaction parameter ${U}$ [$U/t$].}\label{tbl:hubbard}
    \begin{tabular}{l|ccc|ccc} \hline\hline
     \Lower{Method} &
     \mc{3}{c|}{Total energy $[E_h]$} &
     \mc{3}{c}{Correlation energy $[E_h]^\dagger$} \\ 
     & $U/t=1$ & $U/t=2$ & $U/t=3$ & $U/t=1$ & $U/t=2$ & $U/t=3$\\
     \hline \hline
     RHF            &$-$51.203 884&$-$38.703 884&$-$26.203 884&-&-&-\\
    MP2(RHF)$^\star$&$-$52.051 806&$-$42.095 570&$-$33.835 177 &$-$0.847 922&$-$3.391 686&$-$7.631 293\\
MP2(OO-pCCD)$^\bullet$&$-$52.001 992&$-$41.818 037&$-$33.552 933&$-$0.798 108&$-$3.114 153&$-$7.349 049\\
     pCCD           &$-$51.254 320&$-$38.875 387&$-$26.542 450&$-$0.050 435&$-$0.171 503&$-$0.338 566\\
     OO-pCCD        &$-$51.789 439&$-$41.256 975&$-$32.564 839&$-$0.609 173&$-$2.553 095&$-$6.360 955\\
     OO-pCCD-PT2SDd &$-$51.974 875&$-$41.790 323&$-$33.632 419&$-$0.770 991&$-$3.086 443&$-$7.428 535\\
     OO-pCCD-PT2SDo &$-$52.021 235&$-$41.763 875&$-$33.072 400&$-$0.817 351&$-$3.059 991&$-$6.868 516\\
     OO-pCCD-PT2MDd &$-$51.996 174&$-$41.956 922&$-$34.169 938&$-$0.792 290&$-$3.253 038
&$-$7.966 054\\
     OO-pCCD-PT2MDo &$-$52.022 636&$-$41.813 367&$-$33.449 709&$-$0.818 752&$-$3.109 483&$-$7.245 825\\
     OO-pCCD-PT2b   &$-$52.032 318&$-$42.007 466&$-$34.077 901&$-$0.828 434&$-$3.303 582&$-$7.874 017\\
     OO-pCCD-LCCD   &$-$51.994 451&$-$41.745 555&$-$33.195 425 &$-$0.790 567&$-$3.041 671&$-$6.991 541\\
     OO-pCCD-LCCSD  &$-$52.047 516&$-$42.262 870&$-$34.225 963 &$-$0.843 632&$-$3.558 986&$-$8.022 079\\
   CCSD(RHF)$^\star$&$-$52.054 930&$-$42.156 172&$^*$&$-$0.851 046&$-$3.452 288&$^*$\\
CCSD(OO-pCCD)$^\bullet$ &$-$52.054 855&$-$42.164 166&$^*$&$-$0.850 971&$-$3.460 282&$^*$\\ \hline
     Exact$^\ddag$  &$-$52.059 828&$-$42.244 338&$-$34.517 041&$-$0.855 944&$-$3.540 454&$-$8.313 157 \\
      \hline \hline
   \end{tabular} 
\end{center}
}
\begin{tablenotes}\footnotesize
\item $^\dagger$ Calculated w.r.t RHF.\\
\item $^\ddag$ Calculated by solving the Lieb--Wu equations.\\
\item $^\star$ Calculated with an RHF reference function.\\
\item $^\bullet$ Calculated with an OO-pCCD reference function (OO-pCCD orbitals).\\
\item $^*$ Not converged.\\
\end{tablenotes}
\end{table}

\subsection{Electronic structure, orbital entanglement, and electron correlation effects in the model vitamin \ce{B_12} compound}\label{sec:b12-results}
\begin{figure}[tp]                                                
\includegraphics[width=0.5\textwidth]{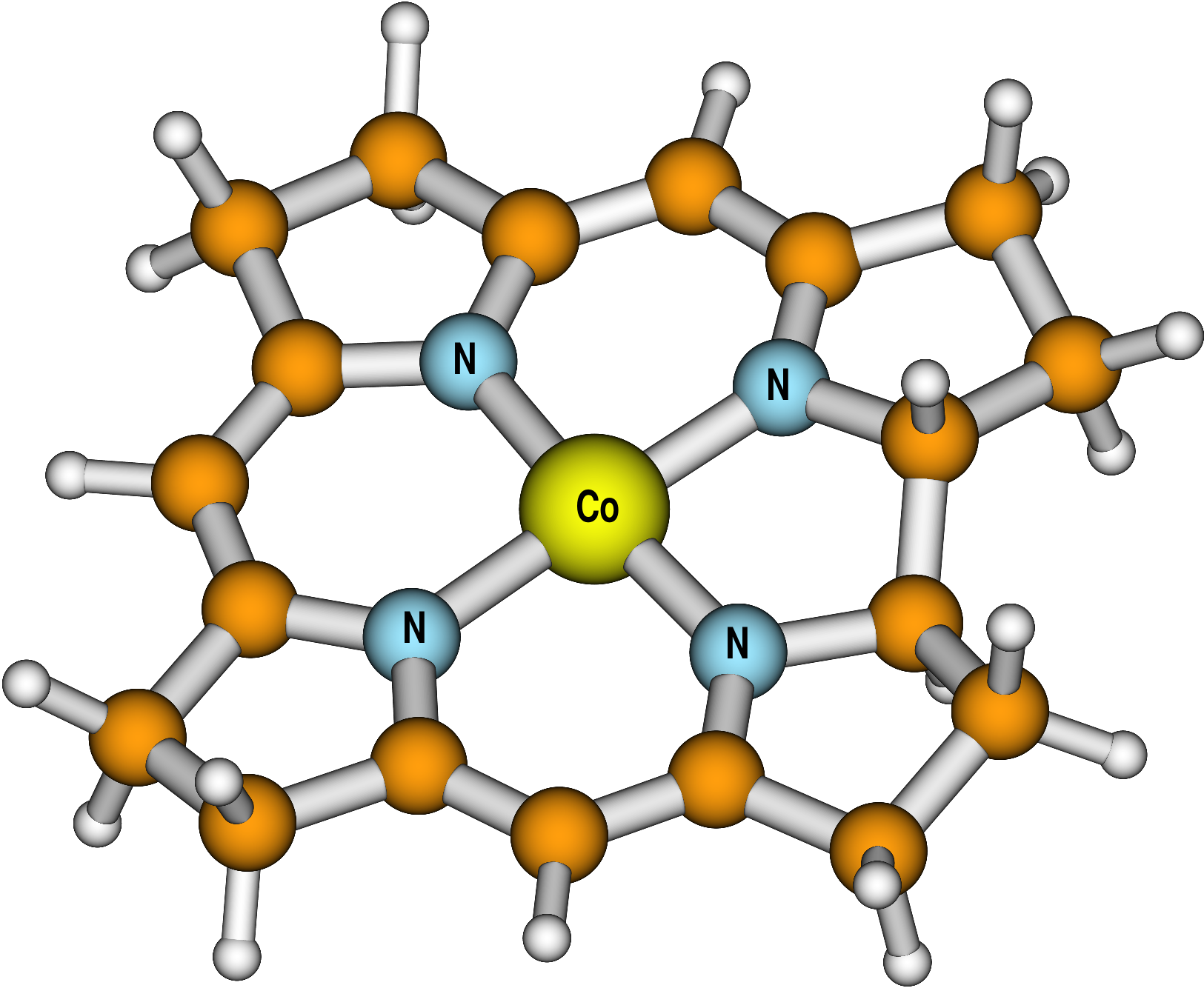}
\caption{The chemical model of the vitamin B$_{12}$ complex that includes the corrin ring. }\label{fig:vitb12label}
\end{figure}

Cobalamins represent a unique group of cobalt-containing complexes that function as cofactors for various enzymes like mutases and transferases.~\cite{ludwig1997structure-vitb12} 
The reactivity of the organometallic Co--C bond, which is broken during catalysis, is an active field of research in bioinorganic chemistry. 
The complex electronic structure and unusually strong multi-reference character of cobalamins attracted a lot of attention from the quantum chemistry community in recent years.~\cite{vitb12-jaworska2003, vitb12-jensen2005,vitb12-liptak2006, vitb12-kumar2011, vitb12-allouche2012, vitb12-kornobis2013, vitb12-kumar2017} 
However, the large size of cobalamin complexes prohibits the application of highly-accurate wavefunction-based quantum chemistry methods and led to the development of cobalt(I)corrin model compounds. 
A simplified, but still realistic model complex is composed of the cobalt atom and a corrin ring (see Figure~\ref{fig:vitb12label}).
This simplified cobalamin-derived compound proved to be an extremely valuable model system in the computational biochemistry community.~\cite{vitb12-jensen2005,vitb12-kumar2017}
The computational challenge with this model compound comes from its complex, yet not fully understood, electronic structure.
A CASSCF/CASPT2 study by Jensen~\cite{vitb12-jensen2005} shows that the ground state electronic structure of cob(I)alamin is composed of a closed-shell Co(I) d$^8$ electron configuration (67\%) and a diradical Co(II)d$^7$-radical corrin $\pi^{*1}$ electron configuration (23\%). 
Yet, the composition of the active space for the cob(I)alamin compound, comprising only 10 electrons distributed in 11 orbitals, CAS(10,11), remains an open question.~\cite{vitb12-jensen2005}   

\begin{figure}[tp]                                                
\includegraphics[width=0.95\textwidth]{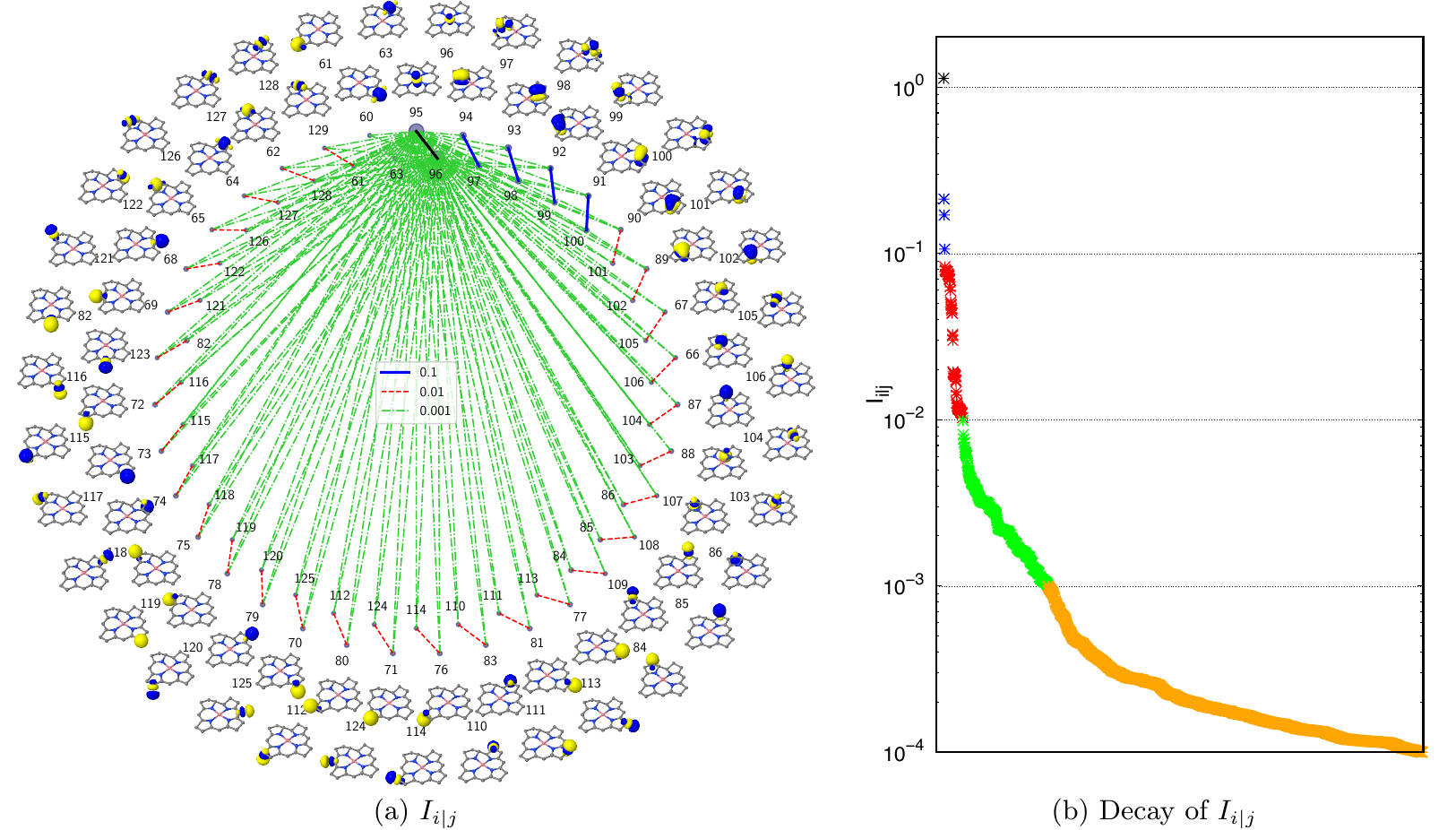}
\caption{Orbital-pair correlation diagrams for the vitamin B$_{12}$ model complex. (a) Orbital-pair mutual information for orbitals nos.~60 to 129. The corresponding isosurface plots are shown in the outermost circle. (b) Decay of the orbital-pair mutual information for $I_{i|j}\ge 10^{-4}$. In total, there are more than 277'000 nonzero orbital-pair correlations. The orbitals were visualized using the \textsc{Jmol} software package.~\cite{jmol}} \label{fig:vitb12-i12}
\end{figure}

This motivated us to study the electronic structure of the model vitamin \ce{B_12} with the OO-pCCD method available in \pybest. 
We used the B3LYP optimized structure from Ref.~\citenum{vitb12-kumar2011} and Dunning's aug-cc-pvdz basis set~\cite{basis_dunning}, which results in a system comprising 190 electrons and 777 orbitals. 
Specifically, in our OO-pCCD calculations, we utilized Cholesky-decomposed electron repulsion integrals (determined with \path{libchol}) with a threshold value of 1e-6 and a frozen core containing the 32 lowest-lying orbitals.
These frozen-core orbitals were optimized by the RHF method.
Thus, our active orbital space comprises 126 electrons distributed in 745 orbitals. 
Our results are summarized in Figure~\ref{fig:vitb12-i12} displaying the orbital-pair mutual information of selected orbitals (nos. 60--129) on the left panel and the decay of the orbital-pair mutual information on the right panel. 
Specifically, the strength of the mutual information is color-coded in descending order, black ($I_{i|j}$> 1.0), blue ($I_{i|j}$> 0.1), red ($I_{i|j}$> 0.001), and green ($I_{i|j}$> 0.001). 
It is evident from Figure~\ref{fig:vitb12-i12}(a) that the largest correlations are between the d$_{z^2}$+$\pi$-type bonding orbital and its d$_{z^2}$+$\pi^*$-type anti-bonding counterpart, and between the $\pi$-type and $\pi^*$-type orbitals of the corrin ring. 
Non-negligible contributions also arise from the $\sigma$-type and $\sigma^*$-type orbitals of the corrin ring. 
That means that all of them (70 in total) should be considered as important in the composition of an active orbital space (in the pCCD optimized basis). 
This is also confirmed by a jump in the decay of $I_{i|j}$ in Figure~\ref{fig:vitb12-i12}(b). 
\section{Conclusions and outlook }\label{sec:conclusions}
We have presented the \pybest software package --- a modern open-source electronic structure platform for \textit{ab inito} calculations at the interface between chemistry and physics. 
One of the strengths of the code is its unique functionality and modern design based on the \texttt{Python3} and \Cpp programming languages interfaced with the \texttt{Pybind11} header-only library. 
In addition to standard quantum chemistry methods, \pybest hosts one-of-a-kind electronic structure approaches based on the pCCD model and its extensions, as well as a quantum entanglement and electron correlation analysis. 
\pybest functionality is demonstrated for the 1-D Hubbard model Hamiltonian and the vitamin \ce{B_12} model system with an active space composed of 126 electrons distributed in 745 orbitals. 

\pybest is designed to host additional software libraries and serves as a hub for future modular software development efforts. 
Specifically, the flexible design of the tensor contraction engine in \pybest, described in section~\ref{sec:implementation}, enables us to conveniently improve bottleneck operations by, for instance, including additional \texttt{Python} or \texttt{numpy} features, interfacing external libraries like \texttt{einsum2},~\cite{einsum2} or optimizing \pybest's internal \texttt{BLAS} routines without the need of changing any wavefunction modules profoundly. 
Moreover, \pybest allows us to exploit the internal parallelization of \texttt{BLAS} and \texttt{numpy.tensordot} (if \texttt{numpy} is linked to a parallel implementation of \texttt{BLAS}). 
Further improvements are possible with the aid of Graphical Processor Units (GPUs) and the \texttt{cupy} array library accelerated by CUDA.~\cite{cupy} 
Besides the technical aspects of code optimization, \pybest also offers an excellent opportunity for the development of novel electronic structure methods. 
Each of the implemented wavefunction modules can be easily extended with additional methods.
A combination of different modules is also possible. 
The latter is particularly useful for the development of the SAPT module, where symmetry adapted perturbation theory can be uniquely combined with pCCD-based methods, as well as for the design and development of embedding-based methods.~\cite{gomes_rev_2012} 
All these features make \pybest an exceptional and flexible programming platform for the quick implementation of unique electronic structure methods, followed by large-scale modelling of electronic structures, their visualization and analysis.
Finally, \pybest is under continuous development and version control (using \textsc{git} on \textsc{Gitlab}), where up to date patches containing bug fixes and improvements are available on the \pybest homepage.~\cite{pybest_web}

\section*{Acknowledgement}
K.B., A.L., and A.N.~acknowledge financial support from a SONATA BIS 5 grant of the National Science Centre, Poland (no.~2015/18/E/ST4/00584). A.L.~acknowledges University of Southern California Research Funding. 
A.N.~received financial support from a PRELUDIUM 17 grant of the National Science Centre, Poland, (no.~2019/33/N/ST4/01880).
P.S.Z. and F.B.~acknowledge financial support from an OPUS 10 grant of the National Science Centre, Poland (no.~2015/19/B/ST4/02707).
P.T.~thanks an OPUS 17 research grant of the National Science Centre, Poland, (no.~2019/33/B/ST4/02114) and a scholarship for outstanding young scientists from the Ministry of Science and Higher Education.

Calculations have been carried out using resources provided by Wroclaw Centre for Networking and Supercomputing (http://wcss.pl), grant nos.~218,~411, and~412. 

We thank Paweł Kozlowski for many helpful discussions concerning the electronic structures of cobalamins. 

\appendix
\section{Example input file for the Hubbard model Hamiltonian}\label{appendix:hubbard}
\begin{lstlisting}[language=Python]
from pybest import *

# define LinalgFactory for 50 sites
lf = DenseLinalgFactory(50)

# define Occupation model and expansion coefficients
occ_model = AufbauOccModel(25)
orb = lf.create_orbital(50)

# initialize Hubbard class
modelham = Hubbard(lf, pbc=True)

# one and two-body interaction terms defined for the on-site basis
# t-param, t = -1
hopping = modelham.compute_one_body(-1)
# U-param, U = 1
onsite = modelham.compute_two_body(1)

# overlap matrix for on-site basis
olp = modelham.compute_overlap()

# do a Hartree-Fock calculation
hf = RHF(lf, occ_model)
hf_ = hf(hopping, onsite, olp, orb)

# do MP2 optimization
mp2 = RMP2(lf, occ_model)
mp2_ = mp2(hopping, onsite, hf_)

# do CCSD optimization
ccsd = RCCSD(lf, occ_model)
ccsd_ = ccsd(hopping, onsite, hf_)

# do OO-pCCD optimization
oopccd = ROOpCCD(lf, occ_model)
oopccd_ = oopccd(hopping, onsite, hf_)
# run until convergence
while(not oopccd_.converged):
    oopccd = ROOpCCD(lf, occ_model)
    oopccd_ = oopccd(hopping, onsite, oopccd_)

# do MP2 optimization with pCCD reference
mp2 = RMP2(lf, occ_model)
mp2_ = mp2(hopping, onsite, oopccd_)

# do CCSD optimization with pCCD reference
ccsd = RCCSD(lf, occ_model)
ccsd_ = ccsd(hopping, onsite, oopccd_)

# do PT2 corrections
pt2 = PT2SDd(lf, occ_model)
pt2_ = pt2(hopping, onsite, oopccd_)

pt2 = PT2SDo(lf, occ_model)
pt2_ = pt2(hopping, onsite, oopccd_)

pt2 = PT2MDd(lf, occ_model)
pt2_ = pt2(hopping, onsite, oopccd_)

pt2 = PT2MDo(lf, occ_model)
pt2_ = pt2(hopping, onsite, oopccd_)

pt2 = PT2b(lf, occ_model)
pt2_ = pt2(hopping, onsite, oopccd_)

# do RpCCD-LCCD calculation
lccd = RpCCDLCCD(lf, occ_model)
lccd_ = lccd(hopping, onsite, oopccd_)

# do RpCCD-LCCSD calculation
lccd = RpCCDLCCSD(lf, occ_model)
lccd_ = lccd(hopping, onsite, oopccd_)

\end{lstlisting}

\section{Example input file for Vit B12}\label{appendix:vitb12}

\begin{lstlisting}[language=Python]
from pybest import *

# set up molecule, define basis set
occ_model = AufbauOccModel(95)
obasis = get_gobasis("aug-cc-pvdz", "coiicor.xyz")

lf = CholeskyLinalgFactory(obasis.nbasis)
orb = lf.create_orbital(obasis.nbasis)
olp = compute_overlap(obasis)

# construct Hamiltonian
kin = compute_kinetic(obasis)
ne = compute_nuclear(obasis)
er = compute_cholesky_eri(obasis, threshold=1e-6)
external = compute_nuclear_repulsion(obasis)

# restricted Hartree-Fock calculation
scf = RHF(lf, occ_model)
hf = scf(kin, ne, er, external, olp, orb)

# orbital-optimized pCCD calculation
pccd = ROOpCCD(lf, occ_model, ncore=32)
pccd_ = pccd(kin, ne, er, hf)

# execute until convergence
while(not pccd_.converged):
    pccd = ROOpCCD(lf, occ_model, ncore=32)
    pccd_ = pccd(kin, ne, er, pccd_)
    # Dump molden file of current solution
    pccd_.gobasis = obasis
    pccd_.to_file("pybest-results/pccd.molden")
    
# perform orbital entanglement and correlation analysis
oe = OrbitalEntanglementRpCCD(lf, pccd_)
oe_= oe()
\end{lstlisting}

The single-orbital entropy and mutual information diagram can be generated using the \path{pybest-entanglement.py} script that is shipped together with the code.
The above example figure can be generated by executing the following command

\begin{lstlisting}[language=Python]
pybest-entanglement.py 0.001 -i 60 -f 129 --sname s1-pccd.dat --iname i12-pccd.dat -z 0.0350 --order 95 96 94 97 93 98 92 99 91 100 90 101 89 102 67 105 66 106 87 104 88 103 107 86 108 85 109 84 77 113 81 111 83 110 76 114 71 124 80 112 70 125 79 120 78 119 75 118 74 117 73 115 72 116 123 82 69 121 68 122 65 126 64 127 62 128 129 61 60 63
\end{lstlisting}
This script features only one positional argument (the threshold for printing the mutual information).
All remaining arguments are optional.
Specifically, the optional argument \path{--order} re-orders the orbitals along the circle and ensures that strongly-correlated orbitals are grouped together (for visualisation purposes only).
\bibliography{rsc}

\providecommand{\latin}[1]{#1}
\makeatletter
\providecommand{\doi}
  {\begingroup\let\do\@makeother\dospecials
  \catcode`\{=1 \catcode`\}=2 \doi@aux}
\providecommand{\doi@aux}[1]{\endgroup\texttt{#1}}
\makeatother
\providecommand*\mcitethebibliography{\thebibliography}
\csname @ifundefined\endcsname{endmcitethebibliography}
  {\let\endmcitethebibliography\endthebibliography}{}
\begin{mcitethebibliography}{88}
\providecommand*\natexlab[1]{#1}
\providecommand*\mciteSetBstSublistMode[1]{}
\providecommand*\mciteSetBstMaxWidthForm[2]{}
\providecommand*\mciteBstWouldAddEndPuncttrue
  {\def\EndOfBibitem{\unskip.}}
\providecommand*\mciteBstWouldAddEndPunctfalse
  {\let\EndOfBibitem\relax}
\providecommand*\mciteSetBstMidEndSepPunct[3]{}
\providecommand*\mciteSetBstSublistLabelBeginEnd[3]{}
\providecommand*\EndOfBibitem{}
\mciteSetBstSublistMode{f}
\mciteSetBstMaxWidthForm{subitem}{(\alph{mcitesubitemcount})}
\mciteSetBstSublistLabelBeginEnd
  {\mcitemaxwidthsubitemform\space}
  {\relax}
  {\relax}

\bibitem[pyb()]{pybind11}
See {\tt https://pybind11.readthedocs.io/en/master/intro.html} for more
  information about the \texttt{pybind11} project (accessed October 1,
  2020)\relax
\mciteBstWouldAddEndPuncttrue
\mciteSetBstMidEndSepPunct{\mcitedefaultmidpunct}
{\mcitedefaultendpunct}{\mcitedefaultseppunct}\relax
\EndOfBibitem
\bibitem[Verstraelen \latin{et~al.}()Verstraelen, Tecmer, Heidar-Zadeh,
  Boguslawski, Chan, Zhao, Kim, Vandenbrande, Yang, Gonzlez-Espinoza,
  \latin{et~al.} others]{horton2.0.0}
Verstraelen,~T.; Tecmer,~P.; Heidar-Zadeh,~F.; Boguslawski,~K.; Chan,~M.;
  Zhao,~Y.; Kim,~T.; Vandenbrande,~S.; Yang,~D.; Gonzlez-Espinoza,~C.~E.,
  \latin{et~al.}  Horton 2.0.0, 2015, {\tt http://theochem.github.com/horton/}
  (accessed September 23, 2020)\relax
\mciteBstWouldAddEndPuncttrue
\mciteSetBstMidEndSepPunct{\mcitedefaultmidpunct}
{\mcitedefaultendpunct}{\mcitedefaultseppunct}\relax
\EndOfBibitem
\bibitem[hor()]{horton3.0.0}
See {\tt https://github.com/theochem/horton} for more information about the
  \textsc{Horton3} project (accessed October 1, 2020)\relax
\mciteBstWouldAddEndPuncttrue
\mciteSetBstMidEndSepPunct{\mcitedefaultmidpunct}
{\mcitedefaultendpunct}{\mcitedefaultseppunct}\relax
\EndOfBibitem
\bibitem[Limacher \latin{et~al.}(2013)Limacher, Ayers, Johnson, {De
  Baerdemacker}, {Van Neck}, and Bultinck]{limacher_2013}
Limacher,~P.~A.; Ayers,~P.~W.; Johnson,~P.~A.; {De Baerdemacker},~S.; {Van
  Neck},~D.; Bultinck,~P. {A New Mean-Field Method Suitable for Strongly
  Correlated Electrons: Computationally Facile Antisymmetric Products of
  Nonorthogonal Geminals}. \emph{J. Chem. Theory Comput.} \textbf{2013},
  \emph{9}, 1394--1401\relax
\mciteBstWouldAddEndPuncttrue
\mciteSetBstMidEndSepPunct{\mcitedefaultmidpunct}
{\mcitedefaultendpunct}{\mcitedefaultseppunct}\relax
\EndOfBibitem
\bibitem[Boguslawski \latin{et~al.}(2014)Boguslawski, Tecmer, Ayers, Bultinck,
  {De Baerdemacker}, and {Van Neck}]{pccd-orbital-optimization}
Boguslawski,~K.; Tecmer,~P.; Ayers,~P.~W.; Bultinck,~P.; {De Baerdemacker},~S.;
  {Van Neck},~D. {Efficient Description Of Strongly Correlated Electrons}.
  \emph{Phys. Rev. B} \textbf{2014}, \emph{89}, 201106(R)\relax
\mciteBstWouldAddEndPuncttrue
\mciteSetBstMidEndSepPunct{\mcitedefaultmidpunct}
{\mcitedefaultendpunct}{\mcitedefaultseppunct}\relax
\EndOfBibitem
\bibitem[Stein \latin{et~al.}(2014)Stein, Henderson, and Scuseria]{tamar-pcc}
Stein,~T.; Henderson,~T.~M.; Scuseria,~G.~E. Seniority Zero Pair Coupled
  Cluster Doubles Theory. \emph{J. Chem. Phys.} \textbf{2014}, \emph{140},
  214113\relax
\mciteBstWouldAddEndPuncttrue
\mciteSetBstMidEndSepPunct{\mcitedefaultmidpunct}
{\mcitedefaultendpunct}{\mcitedefaultseppunct}\relax
\EndOfBibitem
\bibitem[Bartlett and Musia{\l}(2007)Bartlett, and Musia{\l}]{bartlett_2007}
Bartlett,~R.~J.; Musia{\l},~M. Coupled-cluster theory in quantum chemistry.
  \emph{Rev.~Mod.~Phys.} \textbf{2007}, \emph{79}, 291--350\relax
\mciteBstWouldAddEndPuncttrue
\mciteSetBstMidEndSepPunct{\mcitedefaultmidpunct}
{\mcitedefaultendpunct}{\mcitedefaultseppunct}\relax
\EndOfBibitem
\bibitem[Grimme(2003)]{scs-mp2}
Grimme,~S. Improved second-order M\"oller--Plesset perturbation theory by
  separate scaling of parallel- and antiparallel-spin pair correlation
  energies. \emph{J.~Chem.~Phys.} \textbf{2003}, \emph{118}, 9095--9102\relax
\mciteBstWouldAddEndPuncttrue
\mciteSetBstMidEndSepPunct{\mcitedefaultmidpunct}
{\mcitedefaultendpunct}{\mcitedefaultseppunct}\relax
\EndOfBibitem
\bibitem[Rybak \latin{et~al.}(1991)Rybak, Jeziorski, and Szalewicz]{rybak1991}
Rybak,~S.; Jeziorski,~B.; Szalewicz,~K. {Many-Body Symmetry-Adapted
  Perturbation Theory of Intermolecular Interactions. H$_2$O and HF Dimers}.
  \emph{J. Chem. Phys.} \textbf{1991}, \emph{95}, 6576--6601\relax
\mciteBstWouldAddEndPuncttrue
\mciteSetBstMidEndSepPunct{\mcitedefaultmidpunct}
{\mcitedefaultendpunct}{\mcitedefaultseppunct}\relax
\EndOfBibitem
\bibitem[Pulay(1980)]{cdiis}
Pulay,~P. Convergence acceleration of iterative sequences. The case of SCF
  iteration. \emph{Chem.~Phys.~Lett.} \textbf{1980}, \emph{73}, 393--398\relax
\mciteBstWouldAddEndPuncttrue
\mciteSetBstMidEndSepPunct{\mcitedefaultmidpunct}
{\mcitedefaultendpunct}{\mcitedefaultseppunct}\relax
\EndOfBibitem
\bibitem[Kudin \latin{et~al.}(2002)Kudin, Scuseria, and Cancès]{ediis2}
Kudin,~K.~N.; Scuseria,~G.~E.; Cancès,~E. A black-box self-consistent field
  convergence algorithm: One step closer. \emph{J.~Chem.~Phys.} \textbf{2002},
  \emph{116}, 8255--8261\relax
\mciteBstWouldAddEndPuncttrue
\mciteSetBstMidEndSepPunct{\mcitedefaultmidpunct}
{\mcitedefaultendpunct}{\mcitedefaultseppunct}\relax
\EndOfBibitem
\bibitem[Boguslawski \latin{et~al.}(2014)Boguslawski, Tecmer, Ayers, Bultinck,
  {De Baerdemacker}, and {Van Neck}]{pccd-jctc-2014}
Boguslawski,~K.; Tecmer,~P.; Ayers,~P.~W.; Bultinck,~P.; {De Baerdemacker},~S.;
  {Van Neck},~D. {Non-Variational Orbital Optimization Techniques for the
  AP1roG Wave Function}. \emph{J. Chem. Theory Comput.} \textbf{2014},
  \emph{10}, 4873--4882\relax
\mciteBstWouldAddEndPuncttrue
\mciteSetBstMidEndSepPunct{\mcitedefaultmidpunct}
{\mcitedefaultendpunct}{\mcitedefaultseppunct}\relax
\EndOfBibitem
\bibitem[Roos \latin{et~al.}(1980)Roos, Taylor, and Siegbahn]{roos_casscf}
Roos,~B.; Taylor,~P.; Siegbahn,~P. {A complete active space SCF method-(CASSCF)
  using a density matrix formulated super-CI approach}. \emph{Chem. Phys.}
  \textbf{1980}, \emph{48}, 157--173\relax
\mciteBstWouldAddEndPuncttrue
\mciteSetBstMidEndSepPunct{\mcitedefaultmidpunct}
{\mcitedefaultendpunct}{\mcitedefaultseppunct}\relax
\EndOfBibitem
\bibitem[{K. Boguslawski, P. Tecmer}(2015)]{ijqc-entanglement}
{K. Boguslawski, P. Tecmer}, Orbital Entanglement in Quantum Chemistry.
  \emph{Int. J. Quantum Chem.} \textbf{2015}, \emph{115}, 1289--1295\relax
\mciteBstWouldAddEndPuncttrue
\mciteSetBstMidEndSepPunct{\mcitedefaultmidpunct}
{\mcitedefaultendpunct}{\mcitedefaultseppunct}\relax
\EndOfBibitem
\bibitem[{K. Boguslawski, P. Tecmer}(2017)]{ijqc-eratum}
{K. Boguslawski, P. Tecmer}, Erratum: Orbital entanglement in quantum
  chemistry. \emph{Int.~J.~Quantum~Chem.} \textbf{2017}, \emph{117},
  e25455\relax
\mciteBstWouldAddEndPuncttrue
\mciteSetBstMidEndSepPunct{\mcitedefaultmidpunct}
{\mcitedefaultendpunct}{\mcitedefaultseppunct}\relax
\EndOfBibitem
\bibitem[Tecmer \latin{et~al.}(2015)Tecmer, Boguslawski, and
  Ayers]{pawel_pccp2015}
Tecmer,~P.; Boguslawski,~K.; Ayers,~P.~W. Singlet ground state actinide
  chemistry with geminals. \emph{Phys. Chem. Chem. Phys.} \textbf{2015},
  \emph{17}, 14427--14436\relax
\mciteBstWouldAddEndPuncttrue
\mciteSetBstMidEndSepPunct{\mcitedefaultmidpunct}
{\mcitedefaultendpunct}{\mcitedefaultseppunct}\relax
\EndOfBibitem
\bibitem[Boguslawski \latin{et~al.}(2016)Boguslawski, Tecmer, and
  Legeza]{pCCD-prb-2016}
Boguslawski,~K.; Tecmer,~P.; Legeza,~{\"O}. Analysis of two-orbital
  correlations in wavefunctions restricted to electron-pair states.
  \emph{Phys.~Rev.~B} \textbf{2016}, \emph{94}, 155126\relax
\mciteBstWouldAddEndPuncttrue
\mciteSetBstMidEndSepPunct{\mcitedefaultmidpunct}
{\mcitedefaultendpunct}{\mcitedefaultseppunct}\relax
\EndOfBibitem
\bibitem[Boguslawski \latin{et~al.}(2014)Boguslawski, Tecmer, Limacher,
  Johnson, Ayers, Bultinck, {De Baerdemacker}, and {Van Neck}]{ps2-ap1rog}
Boguslawski,~K.; Tecmer,~P.; Limacher,~P.~A.; Johnson,~P.~A.; Ayers,~P.~W.;
  Bultinck,~P.; {De Baerdemacker},~S.; {Van Neck},~D. {Projected Seniority-Two
  Orbital Optimization Of The Antisymmetric Product Of One-Reference Orbital
  Geminal}. \emph{J. Chem. Phys.} \textbf{2014}, \emph{140}, 214114\relax
\mciteBstWouldAddEndPuncttrue
\mciteSetBstMidEndSepPunct{\mcitedefaultmidpunct}
{\mcitedefaultendpunct}{\mcitedefaultseppunct}\relax
\EndOfBibitem
\bibitem[Boguslawski and Tecmer(2017)Boguslawski, and Tecmer]{pccd-PTX}
Boguslawski,~K.; Tecmer,~P. Benchmark of dynamic electron correlation models
  for seniority-zero wavefunctions and their application to thermochemistry.
  \emph{J.~Chem.~Theory~Comput.} \textbf{2017}, \emph{13}, 5966--5983\relax
\mciteBstWouldAddEndPuncttrue
\mciteSetBstMidEndSepPunct{\mcitedefaultmidpunct}
{\mcitedefaultendpunct}{\mcitedefaultseppunct}\relax
\EndOfBibitem
\bibitem[Limacher \latin{et~al.}(2014)Limacher, Ayers, Johnson, {De
  Baerdemacker}, {Van Neck}, and Bultinck]{piotrus_pt2}
Limacher,~P.; Ayers,~P.; Johnson,~P.; {De Baerdemacker},~S.; {Van Neck},~D.;
  Bultinck,~P. {Simple and Inexpensive Perturbative Correction Schemes for
  Antisymmetric Products of Nonorthogonal Geminals}. \emph{Phys. Chem. Chem.
  Phys} \textbf{2014}, \emph{16}, 5061--5065\relax
\mciteBstWouldAddEndPuncttrue
\mciteSetBstMidEndSepPunct{\mcitedefaultmidpunct}
{\mcitedefaultendpunct}{\mcitedefaultseppunct}\relax
\EndOfBibitem
\bibitem[Brz\k{e}k \latin{et~al.}(2019)Brz\k{e}k, Boguslawski, Tecmer, and
  \.{Z}uchowski]{filip-jctc-2019}
Brz\k{e}k,~F.; Boguslawski,~K.; Tecmer,~P.; \.{Z}uchowski,~P.~S. Benchmarking
  the Accuracy of Seniority-Zero Wave Function Methods for Noncovalent
  Interactions. \emph{J.~Chem.~Theory~Comput.} \textbf{2019}, \emph{15},
  4021--4035\relax
\mciteBstWouldAddEndPuncttrue
\mciteSetBstMidEndSepPunct{\mcitedefaultmidpunct}
{\mcitedefaultendpunct}{\mcitedefaultseppunct}\relax
\EndOfBibitem
\bibitem[Boguslawski and Ayers(2015)Boguslawski, and Ayers]{kasia-lcc}
Boguslawski,~K.; Ayers,~P.~W. Linearized Coupled Cluster Correction on the
  Antisymmetric Product of 1-Reference Orbital Geminals.
  \emph{J.~Chem.~Theory~Comput.} \textbf{2015}, \emph{11}, 5252--5261\relax
\mciteBstWouldAddEndPuncttrue
\mciteSetBstMidEndSepPunct{\mcitedefaultmidpunct}
{\mcitedefaultendpunct}{\mcitedefaultseppunct}\relax
\EndOfBibitem
\bibitem[Boguslawski(2016)]{eom-pccd}
Boguslawski,~K. Targeting excited states in all-trans polyenes with
  electron-pair states. \emph{J.~Chem.~Phys.} \textbf{2016}, \emph{145},
  234105\relax
\mciteBstWouldAddEndPuncttrue
\mciteSetBstMidEndSepPunct{\mcitedefaultmidpunct}
{\mcitedefaultendpunct}{\mcitedefaultseppunct}\relax
\EndOfBibitem
\bibitem[Boguslawski(2017)]{eom-pccd-erratum}
Boguslawski,~K. Erratum: “{T}argeting excited states in all-trans polyenes
  with electron-pair states”. \emph{J.~Chem.~Phys.} \textbf{2017},
  \emph{147}, 139901\relax
\mciteBstWouldAddEndPuncttrue
\mciteSetBstMidEndSepPunct{\mcitedefaultmidpunct}
{\mcitedefaultendpunct}{\mcitedefaultseppunct}\relax
\EndOfBibitem
\bibitem[Boguslawski(2019)]{eom-pccd-lccsd}
Boguslawski,~K. Targeting Doubly Excited States with Equation of Motion Coupled
  Cluster Theory Restricted to Double Excitations.
  \emph{J.~Chem.~Theory~Comput.} \textbf{2019}, \emph{15}, 18--24\relax
\mciteBstWouldAddEndPuncttrue
\mciteSetBstMidEndSepPunct{\mcitedefaultmidpunct}
{\mcitedefaultendpunct}{\mcitedefaultseppunct}\relax
\EndOfBibitem
\bibitem[Tecmer \latin{et~al.}(2019)Tecmer, Boguslawski, Borkowski,
  \.{Z}uchowski, and K\k{e}dziera]{pawel-yb2}
Tecmer,~P.; Boguslawski,~K.; Borkowski,~M.; \.{Z}uchowski,~P.~S.;
  K\k{e}dziera,~D. Modeling the electronic structures of the ground and excited
  states of the ytterbium atom and the ytterbium dimer: A modern quantum
  chemistry perspective. \emph{Int.~J.~Quantum~Chem.} \textbf{2019},
  \emph{119}, e25983\relax
\mciteBstWouldAddEndPuncttrue
\mciteSetBstMidEndSepPunct{\mcitedefaultmidpunct}
{\mcitedefaultendpunct}{\mcitedefaultseppunct}\relax
\EndOfBibitem
\bibitem[Nowak \latin{et~al.}(2019)Nowak, Tecmer, and Boguslawski]{Nowak2019}
Nowak,~A.; Tecmer,~P.; Boguslawski,~K. {Assessing the accuracy of simplified
  coupled cluster methods for electronic excited states in f0 actinide
  compounds}. \emph{Phys. Chem. Chem. Phys.} \textbf{2019}, \emph{21},
  19039--19053\relax
\mciteBstWouldAddEndPuncttrue
\mciteSetBstMidEndSepPunct{\mcitedefaultmidpunct}
{\mcitedefaultendpunct}{\mcitedefaultseppunct}\relax
\EndOfBibitem
\bibitem[Brzęk \latin{et~al.}()Brzęk, Leszczyk, Nowak, Boguslawski,
  Kędziera, Tecmer, and Żuchowski]{pybestv1.0.0}
Brzęk,~F.; Leszczyk,~A.; Nowak,~A.; Boguslawski,~K.; Kędziera,~D.;
  Tecmer,~P.; Żuchowski,~P.~S. Pythonic Black-box Electronic Structure Tool
  (PyBEST v1.0.0)\relax
\mciteBstWouldAddEndPuncttrue
\mciteSetBstMidEndSepPunct{\mcitedefaultmidpunct}
{\mcitedefaultendpunct}{\mcitedefaultseppunct}\relax
\EndOfBibitem
\bibitem[pyb()]{pybest_web}
See {\tt http://pybest.fizyka.umk.pl} for more information about
  \textsc{PyBEST} (accessed October 20, 2020)\relax
\mciteBstWouldAddEndPuncttrue
\mciteSetBstMidEndSepPunct{\mcitedefaultmidpunct}
{\mcitedefaultendpunct}{\mcitedefaultseppunct}\relax
\EndOfBibitem
\bibitem[Pipek and Mezey(1989)Pipek, and Mezey]{pipek-mezey}
Pipek,~J.; Mezey,~P.~G. A fast intrinsic localization procedure applicable for
  ab initio and semiempirical linear combination of atomic orbital wave
  functions. \emph{J.~Chem.~Phys.} \textbf{1989}, \emph{90}, 4916--4926\relax
\mciteBstWouldAddEndPuncttrue
\mciteSetBstMidEndSepPunct{\mcitedefaultmidpunct}
{\mcitedefaultendpunct}{\mcitedefaultseppunct}\relax
\EndOfBibitem
\bibitem[{T. Helgaker, P. J{\o}rgensen, J. Olsen}(2000)]{helgaker_book}
{T. Helgaker, P. J{\o}rgensen, J. Olsen}, \emph{Molecular Electronic-Structure
  Theory}; Wiley: New York, 2000\relax
\mciteBstWouldAddEndPuncttrue
\mciteSetBstMidEndSepPunct{\mcitedefaultmidpunct}
{\mcitedefaultendpunct}{\mcitedefaultseppunct}\relax
\EndOfBibitem
\bibitem[{Werner} \latin{et~al.}(2012){Werner}, {Knowles}, Lindh, Manby,
  M.~Sch\"utz, Korona, Mitrushenkov, Rauhut, Adler, Amos, Bernhardsson,
  Berning, Cooper, Deegan, Dobbyn, Eckert, Goll, Hampel, Hetzer, Hrenar,
  Knizia, K\"oppl, Liu, Lloyd, Mata, May, McNicholas, Meyer, Mura, Nicklass,
  Palmieri, Pfl\"uger, Pitzer, Reiher, Schumann, Stoll, Stone, Tarroni,
  Thorsteinsson, Wang, and Wolf]{molpro2012}
{Werner},~H.-J.; {Knowles},~P.~J.; Lindh,~R.; Manby,~F.~R.; M.~Sch\"utz,~P.~C.;
  Korona,~T.; Mitrushenkov,~A.; Rauhut,~G.; Adler,~T.~B.; Amos,~R.~D.;
  Bernhardsson,~A.; Berning,~A.; Cooper,~D.~L.; Deegan,~M. J.~O.;
  Dobbyn,~A.~J.; Eckert,~F.; Goll,~E.; Hampel,~C.; Hetzer,~G.; Hrenar,~T.;
  Knizia,~G.; K\"oppl,~C.; Liu,~Y.; Lloyd,~A.~W.; Mata,~R.~A.; May,~A.~J.;
  McNicholas,~S.~J.; Meyer,~W.; Mura,~M.~E.; Nicklass,~A.; Palmieri,~P.;
  Pfl\"uger,~K.; Pitzer,~R.; Reiher,~M.; Schumann,~U.; Stoll,~H.; Stone,~A.~J.;
  Tarroni,~R.; Thorsteinsson,~T.; Wang,~M.; Wolf,~A. MOLPRO, Version 2012.1, A
  Package Of \emph{Ab Initio} Programs. 2012; see http://www.molpro.net
  (accessed March 1, 2019)\relax
\mciteBstWouldAddEndPuncttrue
\mciteSetBstMidEndSepPunct{\mcitedefaultmidpunct}
{\mcitedefaultendpunct}{\mcitedefaultseppunct}\relax
\EndOfBibitem
\bibitem[Werner \latin{et~al.}(2012)Werner, Knowles, Knizia, Manby, and
  Sch{\"u}tz]{molpro-wires}
Werner,~H.-J.; Knowles,~P.~J.; Knizia,~G.; Manby,~F.~R.; Sch{\"u}tz,~M.
  {Molpro: A General Purpose Quantum Chemistry Program Package}. \emph{WIREs
  Comput. Mol. Sci.} \textbf{2012}, \emph{2}, 242--253\relax
\mciteBstWouldAddEndPuncttrue
\mciteSetBstMidEndSepPunct{\mcitedefaultmidpunct}
{\mcitedefaultendpunct}{\mcitedefaultseppunct}\relax
\EndOfBibitem
\bibitem[Aidas \latin{et~al.}(2014)Aidas, Angeli, Bak, Bakken, Bast, Boman,
  Christiansen, Cimiraglia, Coriani, Dahle, Dalskov, Ekstr{\"{o}}m, Enevoldsen,
  Eriksen, Ettenhuber, Fern{\'{a}}ndez, Ferrighi, Fliegl, Frediani, Hald,
  Halkier, H{\"{a}}ttig, Heiberg, Helgaker, Hennum, Hettema, Hjerten{\ae}s,
  H{\o}st, H{\o}yvik, Iozzi, Jans{\'{i}}k, Jensen, Jonsson, J{\o}rgensen,
  Kauczor, Kirpekar, Kj{\ae}rgaard, Klopper, Knecht, Kobayashi, Koch, Kongsted,
  Krapp, Kristensen, Ligabue, Lutn{\ae}s, Melo, Mikkelsen, Myhre, Neiss,
  Nielsen, Norman, Olsen, Olsen, Osted, Packer, Pawlowski, Pedersen, Provasi,
  Reine, Rinkevicius, Ruden, Ruud, Rybkin, Sa{\l}ek, Samson, de~Mer{\'{a}}s,
  Saue, Sauer, Schimmelpfennig, Sneskov, Steindal, Sylvester-Hvid, Taylor,
  Teale, Tellgren, Tew, Thorvaldsen, Th{\o}gersen, Vahtras, Watson, Wilson,
  Ziolkowski, and {\AA}gren]{dalton-2014}
Aidas,~K.; Angeli,~C.; Bak,~K.~L.; Bakken,~V.; Bast,~R.; Boman,~L.;
  Christiansen,~O.; Cimiraglia,~R.; Coriani,~S.; Dahle,~P.; Dalskov,~E.~K.;
  Ekstr{\"{o}}m,~U.; Enevoldsen,~T.; Eriksen,~J.~J.; Ettenhuber,~P.;
  Fern{\'{a}}ndez,~B.; Ferrighi,~L.; Fliegl,~H.; Frediani,~L.; Hald,~K.;
  Halkier,~A.; H{\"{a}}ttig,~C.; Heiberg,~H.; Helgaker,~T.; Hennum,~A.~C.;
  Hettema,~H.; Hjerten{\ae}s,~E.; H{\o}st,~S.; H{\o}yvik,~I.~M.; Iozzi,~M.~F.;
  Jans{\'{i}}k,~B.; Jensen,~H. J.~A.; Jonsson,~D.; J{\o}rgensen,~P.;
  Kauczor,~J.; Kirpekar,~S.; Kj{\ae}rgaard,~T.; Klopper,~W.; Knecht,~S.;
  Kobayashi,~R.; Koch,~H.; Kongsted,~J.; Krapp,~A.; Kristensen,~K.;
  Ligabue,~A.; Lutn{\ae}s,~O.~B.; Melo,~J.~I.; Mikkelsen,~K.~V.; Myhre,~R.~H.;
  Neiss,~C.; Nielsen,~C.~B.; Norman,~P.; Olsen,~J.; Olsen,~J. M.~H.; Osted,~A.;
  Packer,~M.~J.; Pawlowski,~F.; Pedersen,~T.~B.; Provasi,~P.~F.; Reine,~S.;
  Rinkevicius,~Z.; Ruden,~T.~A.; Ruud,~K.; Rybkin,~V.~V.; Sa{\l}ek,~P.;
  Samson,~C.~C.; de~Mer{\'{a}}s,~A.~S.; Saue,~T.; Sauer,~S.~P.;
  Schimmelpfennig,~B.; Sneskov,~K.; Steindal,~A.~H.; Sylvester-Hvid,~K.~O.;
  Taylor,~P.~R.; Teale,~A.~M.; Tellgren,~E.~I.; Tew,~D.~P.; Thorvaldsen,~A.~J.;
  Th{\o}gersen,~L.; Vahtras,~O.; Watson,~M.~A.; Wilson,~D.~J.; Ziolkowski,~M.;
  {\AA}gren,~H. {The Dalton quantum chemistry program system}. \emph{WIREs
  Comput. Mol. Sci.} \textbf{2014}, \emph{4}, 269--284\relax
\mciteBstWouldAddEndPuncttrue
\mciteSetBstMidEndSepPunct{\mcitedefaultmidpunct}
{\mcitedefaultendpunct}{\mcitedefaultseppunct}\relax
\EndOfBibitem
\bibitem[Legeza()]{dmrg_ors}
Legeza,~{\"O}. \textsc{QC-DMRG-Budapest}, A Program for Quantum Chemical {DMRG}
  Calculations. { \rm Copyright 2000--2020, HAS RISSPO Budapest}\relax
\mciteBstWouldAddEndPuncttrue
\mciteSetBstMidEndSepPunct{\mcitedefaultmidpunct}
{\mcitedefaultendpunct}{\mcitedefaultseppunct}\relax
\EndOfBibitem
\bibitem[Hubbard(1963)]{hubbard_model}
Hubbard,~J. {Electron Correlations in Narrow Energy Bands}. \emph{Proc. R. Soc.
  Lond. A} \textbf{1963}, \emph{276}, 238--257\relax
\mciteBstWouldAddEndPuncttrue
\mciteSetBstMidEndSepPunct{\mcitedefaultmidpunct}
{\mcitedefaultendpunct}{\mcitedefaultseppunct}\relax
\EndOfBibitem
\bibitem[Douglas and Kroll(1974)Douglas, and Kroll]{dkh1}
Douglas,~N.; Kroll,~N.~M. Quantum Electrodynamical Corrections To
  Fine-Structure Of Helium. \emph{Ann.~Phys.} \textbf{1974}, \emph{82},
  89--155\relax
\mciteBstWouldAddEndPuncttrue
\mciteSetBstMidEndSepPunct{\mcitedefaultmidpunct}
{\mcitedefaultendpunct}{\mcitedefaultseppunct}\relax
\EndOfBibitem
\bibitem[Hess(1986)]{dkh2}
Hess,~B.~A. Relativistic Electronic-Structure Calculations Employing A
  2-Component No-Pair Formalism With External-Fields Projection Operators.
  \emph{Phys.~Rev.~A} \textbf{1986}, \emph{33}, 3742--3748\relax
\mciteBstWouldAddEndPuncttrue
\mciteSetBstMidEndSepPunct{\mcitedefaultmidpunct}
{\mcitedefaultendpunct}{\mcitedefaultseppunct}\relax
\EndOfBibitem
\bibitem[K\k{e}dziera(2005)]{darek-dirac-limit}
K\k{e}dziera,~D. Convergence of approximate two-component Hamiltonians: How far
  is the Dirac limit. \emph{J.~Chem.~Phys.} \textbf{2005}, \emph{123},
  074109\relax
\mciteBstWouldAddEndPuncttrue
\mciteSetBstMidEndSepPunct{\mcitedefaultmidpunct}
{\mcitedefaultendpunct}{\mcitedefaultseppunct}\relax
\EndOfBibitem
\bibitem[Reiher and Wolf(2009)Reiher, and Wolf]{reiher_book}
Reiher,~M.; Wolf,~A. \emph{Relativistic Quantum Chemistry. {T}he Fundamental
  Theory of Molecular Science}; Wiley, 2009\relax
\mciteBstWouldAddEndPuncttrue
\mciteSetBstMidEndSepPunct{\mcitedefaultmidpunct}
{\mcitedefaultendpunct}{\mcitedefaultseppunct}\relax
\EndOfBibitem
\bibitem[Reiher(2012)]{reiher2012relativistic}
Reiher,~M. Relativistic Douglas--Kroll--Hess theory.
  \emph{Wiley~Interdiscip.~Rev.~Comput.~Mol.~Sci.} \textbf{2012}, \emph{2},
  139--149\relax
\mciteBstWouldAddEndPuncttrue
\mciteSetBstMidEndSepPunct{\mcitedefaultmidpunct}
{\mcitedefaultendpunct}{\mcitedefaultseppunct}\relax
\EndOfBibitem
\bibitem[{P. Tecmer, K. Boguslawski, D.
  K{\c{e}}dziera}(2017)]{pawel-relativistic-book-chapter-2016}
{P. Tecmer, K. Boguslawski, D. K{\c{e}}dziera}, In \emph{Handbook of
  Computational Chemistry}; Leszczy{\'{n}}ski,~J., Ed.; Springer Netherlands:
  Dordrecht, 2017; Vol.~2; pp 885--926\relax
\mciteBstWouldAddEndPuncttrue
\mciteSetBstMidEndSepPunct{\mcitedefaultmidpunct}
{\mcitedefaultendpunct}{\mcitedefaultseppunct}\relax
\EndOfBibitem
\bibitem[Aquilante \latin{et~al.}(2011)Aquilante, Boman, Bostr{\"o}m, Koch,
  Lindh, de~Mer{\'a}s, and Pedersen]{cholesky-review-2011}
Aquilante,~F.; Boman,~L.; Bostr{\"o}m,~J.; Koch,~H.; Lindh,~R.;
  de~Mer{\'a}s,~A.~S.; Pedersen,~T.~B. \emph{Linear-Scaling Techniques in
  Computational Chemistry and Physics}; Springer, 2011; pp 301--343\relax
\mciteBstWouldAddEndPuncttrue
\mciteSetBstMidEndSepPunct{\mcitedefaultmidpunct}
{\mcitedefaultendpunct}{\mcitedefaultseppunct}\relax
\EndOfBibitem
\bibitem[Rabuck and Scuseria(1999)Rabuck, and Scuseria]{fermi-occ-model-99}
Rabuck,~A.~D.; Scuseria,~G.~E. Improving self-consistent field convergence by
  varying occupation numbers. \emph{J.~Chem.~Phys.} \textbf{1999}, \emph{110},
  695--700\relax
\mciteBstWouldAddEndPuncttrue
\mciteSetBstMidEndSepPunct{\mcitedefaultmidpunct}
{\mcitedefaultendpunct}{\mcitedefaultseppunct}\relax
\EndOfBibitem
\bibitem[Weinhold and Wilson(1967)Weinhold, and Wilson]{doci}
Weinhold,~F.; Wilson,~E.~B. {Reduced Density Matrices Of Atoms And Molecules.
  I. The 2 Matrix Of Double-Occupancy, Configuration-Interaction Wavefunctions
  For Singlet States}. \emph{J. Chem. Phys.} \textbf{1967}, \emph{46},
  2752--2758\relax
\mciteBstWouldAddEndPuncttrue
\mciteSetBstMidEndSepPunct{\mcitedefaultmidpunct}
{\mcitedefaultendpunct}{\mcitedefaultseppunct}\relax
\EndOfBibitem
\bibitem[Dennis and Mei(1979)Dennis, and Mei]{double-dogleg}
Dennis,~J.~E.; Mei,~H. Two new unconstrained optimization algorithms which use
  function and gradient values. \emph{J.~Optim. Theory Applications}
  \textbf{1979}, \emph{28}, 453--482\relax
\mciteBstWouldAddEndPuncttrue
\mciteSetBstMidEndSepPunct{\mcitedefaultmidpunct}
{\mcitedefaultendpunct}{\mcitedefaultseppunct}\relax
\EndOfBibitem
\bibitem[Steihaug(1983)]{steihaug1983conjugate}
Steihaug,~T. The conjugate gradient method and trust regions in large scale
  optimization. \emph{SIAM J.~Num.~Anal.} \textbf{1983}, \emph{20},
  626--637\relax
\mciteBstWouldAddEndPuncttrue
\mciteSetBstMidEndSepPunct{\mcitedefaultmidpunct}
{\mcitedefaultendpunct}{\mcitedefaultseppunct}\relax
\EndOfBibitem
\bibitem[Powell(1970)]{powell1970}
Powell,~M. In \emph{Nonlinear Programming}; Rosen,~J., Mangasarian,~O.,
  Ritter,~K., Eds.; Academic Press, New York, 1970; pp 31--65\relax
\mciteBstWouldAddEndPuncttrue
\mciteSetBstMidEndSepPunct{\mcitedefaultmidpunct}
{\mcitedefaultendpunct}{\mcitedefaultseppunct}\relax
\EndOfBibitem
\bibitem[Tecmer \latin{et~al.}(2014)Tecmer, Boguslawski, Limacher, Johnson,
  Chan, Verstraelen, and Ayers]{pawel_jpca_2014}
Tecmer,~P.; Boguslawski,~K.; Limacher,~P.~A.; Johnson,~P.~A.; Chan,~M.;
  Verstraelen,~T.; Ayers,~P.~W. {Assessing the Accuracy of New Geminal-Based
  Approaches}. \emph{J. Phys. Chem. A} \textbf{2014}, \emph{118},
  9058--9068\relax
\mciteBstWouldAddEndPuncttrue
\mciteSetBstMidEndSepPunct{\mcitedefaultmidpunct}
{\mcitedefaultendpunct}{\mcitedefaultseppunct}\relax
\EndOfBibitem
\bibitem[Leszczyk \latin{et~al.}(2019)Leszczyk, Tecmer, and
  Boguslawski]{ola-book-chapter-2019}
Leszczyk,~A.; Tecmer,~P.; Boguslawski,~K. \emph{{Transition Metals in
  Coordination Environments, Challenges and Advances in Computational Chemistry
  and Physics}}; Springer: Cham (Switzerland), 2019; Vol.~29; Chapter {New
  Strategies in Modeling Electronic Structures and Properties with Applications
  to Actinides}, pp 121--160\relax
\mciteBstWouldAddEndPuncttrue
\mciteSetBstMidEndSepPunct{\mcitedefaultmidpunct}
{\mcitedefaultendpunct}{\mcitedefaultseppunct}\relax
\EndOfBibitem
\bibitem[Virtanen \latin{et~al.}(2020)Virtanen, Gommers, Oliphant, Haberland,
  Reddy, Cournapeau, Burovski, Peterson, Weckesser, Bright, \latin{et~al.}
  others]{scipy1.0}
Virtanen,~P.; Gommers,~R.; Oliphant,~T.~E.; Haberland,~M.; Reddy,~T.;
  Cournapeau,~D.; Burovski,~E.; Peterson,~P.; Weckesser,~W.; Bright,~J.,
  \latin{et~al.}  SciPy 1.0: fundamental algorithms for scientific computing in
  Python. \emph{Nat. Methods} \textbf{2020}, \emph{17}, 261--272\relax
\mciteBstWouldAddEndPuncttrue
\mciteSetBstMidEndSepPunct{\mcitedefaultmidpunct}
{\mcitedefaultendpunct}{\mcitedefaultseppunct}\relax
\EndOfBibitem
\bibitem[Nowak \latin{et~al.}(2020)Nowak, Legeza, and
  Boguslawski]{artur-entanglement-pccd-lccsd}
Nowak,~A.; Legeza,~O.; Boguslawski,~K. Orbital entanglement and correlation
  from pCCD-tailored Coupled Cluster wave functions. \emph{arXiv}
  \textbf{2020}, arXiv:2010.01934\relax
\mciteBstWouldAddEndPuncttrue
\mciteSetBstMidEndSepPunct{\mcitedefaultmidpunct}
{\mcitedefaultendpunct}{\mcitedefaultseppunct}\relax
\EndOfBibitem
\bibitem[Henderson \latin{et~al.}(2014)Henderson, Bulik, Stein, and
  Scuseria]{frozen-pccd}
Henderson,~T.~M.; Bulik,~I.~W.; Stein,~T.; Scuseria,~G.~E. Seniority-based
  coupled cluster theory. \emph{J.~Chem.~Phys.} \textbf{2014}, \emph{141},
  244104\relax
\mciteBstWouldAddEndPuncttrue
\mciteSetBstMidEndSepPunct{\mcitedefaultmidpunct}
{\mcitedefaultendpunct}{\mcitedefaultseppunct}\relax
\EndOfBibitem
\bibitem[Veis \latin{et~al.}(2016)Veis, Antal{\'i}k, Brabec, Neese, Legeza, and
  Pittner]{veis2016}
Veis,~L.; Antal{\'i}k,~A.; Brabec,~J.; Neese,~F.; Legeza,~{\"O}.; Pittner,~J.
  Coupled Cluster Method with Single and Double Excitations Tailored by Matrix
  Product State Wave Functions. \emph{J. Phys. Chem. Lett.} \textbf{2016},
  \emph{7}, 4072--4078\relax
\mciteBstWouldAddEndPuncttrue
\mciteSetBstMidEndSepPunct{\mcitedefaultmidpunct}
{\mcitedefaultendpunct}{\mcitedefaultseppunct}\relax
\EndOfBibitem
\bibitem[Jeziorski \latin{et~al.}(1994)Jeziorski, Moszy\'n{}ski, and
  Szalewicz]{jeziorski1994}
Jeziorski,~B.; Moszy\'n{}ski,~R.; Szalewicz,~K. {Perturbation Theory Approach
  to Intermolecular Potential Energy Surfaces of van der Waals Complexes}.
  \emph{Chem. Rev.} \textbf{1994}, \emph{94}, 1887--1930\relax
\mciteBstWouldAddEndPuncttrue
\mciteSetBstMidEndSepPunct{\mcitedefaultmidpunct}
{\mcitedefaultendpunct}{\mcitedefaultseppunct}\relax
\EndOfBibitem
\bibitem[Patkowski(2020)]{patkowski-sapt-review-2020}
Patkowski,~K. Recent developments in symmetry-adapted perturbation theory.
  \emph{Wiley~Interdiscip.~Rev.~Comput.~Mol.~Sci.} \textbf{2020}, \emph{10},
  e1452\relax
\mciteBstWouldAddEndPuncttrue
\mciteSetBstMidEndSepPunct{\mcitedefaultmidpunct}
{\mcitedefaultendpunct}{\mcitedefaultseppunct}\relax
\EndOfBibitem
\bibitem[Rissler \latin{et~al.}(2006)Rissler, Noack, and White]{rissler2006}
Rissler,~J.; Noack,~R.~M.; White,~S.~R. {Measuring Orbital Interaction Using
  Quantum Information Theory}. \emph{Chem. Phys.} \textbf{2006}, \emph{323},
  519--531\relax
\mciteBstWouldAddEndPuncttrue
\mciteSetBstMidEndSepPunct{\mcitedefaultmidpunct}
{\mcitedefaultendpunct}{\mcitedefaultseppunct}\relax
\EndOfBibitem
\bibitem[Barcza \latin{et~al.}(2014)Barcza, Noack, S{\'o}lyom, and
  Legeza]{barcza2014entanglement}
Barcza,~G.; Noack,~R.; S{\'o}lyom,~J.; Legeza,~{\"O}. Entanglement patterns and
  generalized correlation functions in quantum many body systems.
  \emph{Phys.~Rev.~B} \textbf{2014}, \emph{92}, 125140\relax
\mciteBstWouldAddEndPuncttrue
\mciteSetBstMidEndSepPunct{\mcitedefaultmidpunct}
{\mcitedefaultendpunct}{\mcitedefaultseppunct}\relax
\EndOfBibitem
\bibitem[Boguslawski \latin{et~al.}(2012)Boguslawski, Tecmer, Legeza, and
  Reiher]{entanglement_letter}
Boguslawski,~K.; Tecmer,~P.; Legeza,~O.; Reiher,~M. {Entanglement Measures for
  Single- and Multireference Correlation Effects}. \emph{J. Phys. Chem. Lett.}
  \textbf{2012}, \emph{3}, 3129--3135\relax
\mciteBstWouldAddEndPuncttrue
\mciteSetBstMidEndSepPunct{\mcitedefaultmidpunct}
{\mcitedefaultendpunct}{\mcitedefaultseppunct}\relax
\EndOfBibitem
\bibitem[Tecmer \latin{et~al.}(2014)Tecmer, Boguslawski, Legeza, and
  Reiher]{cuo_dmrg}
Tecmer,~P.; Boguslawski,~K.; Legeza,~O.; Reiher,~M. {Unravelling the
  Quantum-Entanglement Effect of Noble Gas Coordination on the Spin Ground
  State of CUO}. \emph{Phys. Chem. Chem. Phys} \textbf{2014}, \emph{16},
  719--727\relax
\mciteBstWouldAddEndPuncttrue
\mciteSetBstMidEndSepPunct{\mcitedefaultmidpunct}
{\mcitedefaultendpunct}{\mcitedefaultseppunct}\relax
\EndOfBibitem
\bibitem[Duperrouzel \latin{et~al.}(2015)Duperrouzel, Tecmer, Boguslawski,
  Barcza, Legeza, and Ayers]{corinne_2015}
Duperrouzel,~C.; Tecmer,~P.; Boguslawski,~K.; Barcza,~G.; Legeza,~O.;
  Ayers,~P.~W. {A Quantum Informational Approach for Dissecting Chemical
  Reactions}. \emph{Chem. Phys. Lett.} \textbf{2015}, \emph{621},
  160--164\relax
\mciteBstWouldAddEndPuncttrue
\mciteSetBstMidEndSepPunct{\mcitedefaultmidpunct}
{\mcitedefaultendpunct}{\mcitedefaultseppunct}\relax
\EndOfBibitem
\bibitem[Freitag \latin{et~al.}(2015)Freitag, Knecht, Keller, Delcey,
  Aquilante, Pedersen, Lindh, Reiher, and Gonzalez]{roland-runo}
Freitag,~L.; Knecht,~S.; Keller,~S.~F.; Delcey,~M.~G.; Aquilante,~F.;
  Pedersen,~T.~B.; Lindh,~R.; Reiher,~M.; Gonzalez,~L. Orbital entanglement and
  CASSCF analysis of the Ru-NO bond in a Ruthenium nitrosyl complex.
  \emph{Phys. Chem. Chem. Phys.} \textbf{2015}, \emph{17}, 13769--13769\relax
\mciteBstWouldAddEndPuncttrue
\mciteSetBstMidEndSepPunct{\mcitedefaultmidpunct}
{\mcitedefaultendpunct}{\mcitedefaultseppunct}\relax
\EndOfBibitem
\bibitem[Stein and Reiher(2016)Stein, and Reiher]{stein2016}
Stein,~C.~J.; Reiher,~M. {Automated Selection of Active Orbital Spaces}.
  \emph{J.~Chem.~Theory~Comput.} \textbf{2016}, \emph{12}, 1760--1771\relax
\mciteBstWouldAddEndPuncttrue
\mciteSetBstMidEndSepPunct{\mcitedefaultmidpunct}
{\mcitedefaultendpunct}{\mcitedefaultseppunct}\relax
\EndOfBibitem
\bibitem[Boguslawski \latin{et~al.}(2017)Boguslawski, R\'{e}al, Tecmer,
  Duperrouzel, Gomes, Legeza, Ayers, and
  Vallet]{boguslawski-plutonium-oxides-2017}
Boguslawski,~K.; R\'{e}al,~F.; Tecmer,~P.; Duperrouzel,~C.; Gomes,~A. S.~P.;
  Legeza,~{\"O}.; Ayers,~P.~W.; Vallet,~V. On the multi-reference nature of
  plutonium oxides: {PuO$_2^{2+}$}, {PuO$_2$}, {PuO$_3$} and {PuO$_2$(OH)$_2$}.
  \emph{Phys. Chem. Chem. Phys.} \textbf{2017}, \emph{19}, 4317--4329\relax
\mciteBstWouldAddEndPuncttrue
\mciteSetBstMidEndSepPunct{\mcitedefaultmidpunct}
{\mcitedefaultendpunct}{\mcitedefaultseppunct}\relax
\EndOfBibitem
\bibitem[\L{}achma\'{n}ska \latin{et~al.}(2019)\L{}achma\'{n}ska, Tecmer,
  Legeza, and Boguslawski]{ola-pccp-2019}
\L{}achma\'{n}ska,~A.; Tecmer,~P.; Legeza,~{\"O}.; Boguslawski,~K. Elucidating
  cation–cation interactions in neptunyl dications using multi-reference ab
  initio theory. \emph{Phys. Chem. Chem. Phys.} \textbf{2019}, \emph{21},
  744--759\relax
\mciteBstWouldAddEndPuncttrue
\mciteSetBstMidEndSepPunct{\mcitedefaultmidpunct}
{\mcitedefaultendpunct}{\mcitedefaultseppunct}\relax
\EndOfBibitem
\bibitem[lib()]{libint}
A library for the evaluation of molecular integrals of many-body operators over
  Gaussian functions E. F. Valeev; (2019), {\tt http://libint.valeyev.net/}
  (accessed October 1, 2020)\relax
\mciteBstWouldAddEndPuncttrue
\mciteSetBstMidEndSepPunct{\mcitedefaultmidpunct}
{\mcitedefaultendpunct}{\mcitedefaultseppunct}\relax
\EndOfBibitem
\bibitem[mkl()]{mkl}
See {\tt
  https://software.intel.com/content/www/us/en/develop/tools/math-kernel-library.html}
  for more information about Intel MKL (accessed September 24, 2020)\relax
\mciteBstWouldAddEndPuncttrue
\mciteSetBstMidEndSepPunct{\mcitedefaultmidpunct}
{\mcitedefaultendpunct}{\mcitedefaultseppunct}\relax
\EndOfBibitem
\bibitem[Smith and Gray(2018)Smith, and Gray]{opt_einsum-2018}
Smith,~D. G.~A.; Gray,~J. opt\_einsum - A Python package for optimizing
  contraction order for einsum-like expressions. \emph{J. Open Source Softw.}
  \textbf{2018}, \emph{3}, 753\relax
\mciteBstWouldAddEndPuncttrue
\mciteSetBstMidEndSepPunct{\mcitedefaultmidpunct}
{\mcitedefaultendpunct}{\mcitedefaultseppunct}\relax
\EndOfBibitem
\bibitem[Harris \latin{et~al.}(2020)Harris, Millman, {van der Walt}, Gommers,
  Virtanen, Cournapeau, Wieser, Taylor, Berg, Smith, Kern, Picus, Hoyer, {van
  Kerkwijk}, Brett, Haldane, {del R\'{i}o}, Wiebe, Peterson,
  G\'{e}rard-Marchant, Sheppard, Reddy, Weckesser, Abbasi, Gohlke, and
  Oliphant]{numpy-2020}
Harris,~C.~R.; Millman,~K.~J.; {van der Walt},~S.~J.; Gommers,~R.;
  Virtanen,~P.; Cournapeau,~D.; Wieser,~E.; Taylor,~J.; Berg,~S.; Smith,~N.~J.;
  Kern,~R.; Picus,~M.; Hoyer,~S.; {van Kerkwijk},~M.~H.; Brett,~M.;
  Haldane,~A.; {del R\'{i}o},~J.~F.; Wiebe,~M.; Peterson,~P.;
  G\'{e}rard-Marchant,~P.; Sheppard,~K.; Reddy,~T.; Weckesser,~W.; Abbasi,~H.;
  Gohlke,~C.; Oliphant,~T.~E. Array programming with NumPy. \emph{Nature}
  \textbf{2020}, \emph{585}, 357--362\relax
\mciteBstWouldAddEndPuncttrue
\mciteSetBstMidEndSepPunct{\mcitedefaultmidpunct}
{\mcitedefaultendpunct}{\mcitedefaultseppunct}\relax
\EndOfBibitem
\bibitem[Smith \latin{et~al.}(2018)Smith, Burns, Sirianni, Nascimento, Kumar,
  James, Schriber, Zhang, Zhang, Abbott, \latin{et~al.} others]{psi4numpy}
Smith,~D.~G.; Burns,~L.~A.; Sirianni,~D.~A.; Nascimento,~D.~R.; Kumar,~A.;
  James,~A.~M.; Schriber,~J.~B.; Zhang,~T.; Zhang,~B.; Abbott,~A.~S.,
  \latin{et~al.}  Psi4NumPy: An interactive quantum chemistry programming
  environment for reference implementations and rapid development.
  \emph{J.~Chem.~Theory~Comput.} \textbf{2018}, \emph{14}, 3504--3511\relax
\mciteBstWouldAddEndPuncttrue
\mciteSetBstMidEndSepPunct{\mcitedefaultmidpunct}
{\mcitedefaultendpunct}{\mcitedefaultseppunct}\relax
\EndOfBibitem
\bibitem[sph()]{sphinx}
See {\tt https://www.sphinx-doc.org/en/master/} for more information about
  \texttt{Sphinx} (accessed October 1, 2020)\relax
\mciteBstWouldAddEndPuncttrue
\mciteSetBstMidEndSepPunct{\mcitedefaultmidpunct}
{\mcitedefaultendpunct}{\mcitedefaultseppunct}\relax
\EndOfBibitem
\bibitem[zen()]{zenodo}
See {\tt https://zenodo.org} for more information about the \textsc{Zenodo}
  plaftorm (accessed October 1, 2020)\relax
\mciteBstWouldAddEndPuncttrue
\mciteSetBstMidEndSepPunct{\mcitedefaultmidpunct}
{\mcitedefaultendpunct}{\mcitedefaultseppunct}\relax
\EndOfBibitem
\bibitem[Rodr\'{\i}guez-Guzm\'{a}n
  \latin{et~al.}(2013)Rodr\'{\i}guez-Guzm\'{a}n, Jim\'{e}nez-Hoyos, Schutski,
  and Scuseria]{hubbard-gustavo_2013}
Rodr\'{\i}guez-Guzm\'{a}n,~R.; Jim\'{e}nez-Hoyos,~C.~A.; Schutski,~R.;
  Scuseria,~G.~E. {Multireference Symmetry-Projected Variational Approaches for
  Ground and Excited States of the One-Dimensional Hubbard Model}. \emph{Phys.
  Rev. B} \textbf{2013}, \emph{87}, 235129\relax
\mciteBstWouldAddEndPuncttrue
\mciteSetBstMidEndSepPunct{\mcitedefaultmidpunct}
{\mcitedefaultendpunct}{\mcitedefaultseppunct}\relax
\EndOfBibitem
\bibitem[Lieb and Wu(1968)Lieb, and Wu]{lieb-wu}
Lieb,~E.~H.; Wu,~F.~Y. Absence of Mott Transition in an Exact Solution of the
  Short-Range, One-Band Model in One Dimension. \emph{Phys. Rev. Lett.}
  \textbf{1968}, \emph{20}, 1445--1448\relax
\mciteBstWouldAddEndPuncttrue
\mciteSetBstMidEndSepPunct{\mcitedefaultmidpunct}
{\mcitedefaultendpunct}{\mcitedefaultseppunct}\relax
\EndOfBibitem
\bibitem[Ludwig and Matthews(1997)Ludwig, and
  Matthews]{ludwig1997structure-vitb12}
Ludwig,~M.~L.; Matthews,~R.~G. Structure-based perspectives on
  \ce{B_12}-dependent enzymes. \emph{Ann. Rev. Biochem.} \textbf{1997},
  \emph{66}, 269--313\relax
\mciteBstWouldAddEndPuncttrue
\mciteSetBstMidEndSepPunct{\mcitedefaultmidpunct}
{\mcitedefaultendpunct}{\mcitedefaultseppunct}\relax
\EndOfBibitem
\bibitem[Jaworska and Lodowski(2003)Jaworska, and
  Lodowski]{vitb12-jaworska2003}
Jaworska,~M.; Lodowski,~P. {Electronic spectrum of Co-corrin calculated with
  the TDDFT method}. \emph{J. Mol. Struc. (Theochem)} \textbf{2003},
  \emph{631}, 209--223\relax
\mciteBstWouldAddEndPuncttrue
\mciteSetBstMidEndSepPunct{\mcitedefaultmidpunct}
{\mcitedefaultendpunct}{\mcitedefaultseppunct}\relax
\EndOfBibitem
\bibitem[Jensen(2005)]{vitb12-jensen2005}
Jensen,~K.~P. Electronic structure of Cob(I)alamin: the story of an unusual
  nucleophile. \emph{J.~Phys.~Chem.~B} \textbf{2005}, \emph{109},
  10505--10512\relax
\mciteBstWouldAddEndPuncttrue
\mciteSetBstMidEndSepPunct{\mcitedefaultmidpunct}
{\mcitedefaultendpunct}{\mcitedefaultseppunct}\relax
\EndOfBibitem
\bibitem[Liptak and Brunold(2006)Liptak, and Brunold]{vitb12-liptak2006}
Liptak,~M.~D.; Brunold,~T.~C. Spectroscopic and computational studies of
  \ce{Co^{1+}} cobalamin: spectral and electronic properties of the
  “superreduced” \ce{B_12} cofactor. \emph{J.~Am.~Chem.~Soc.}
  \textbf{2006}, \emph{128}, 9144--9156\relax
\mciteBstWouldAddEndPuncttrue
\mciteSetBstMidEndSepPunct{\mcitedefaultmidpunct}
{\mcitedefaultendpunct}{\mcitedefaultseppunct}\relax
\EndOfBibitem
\bibitem[Kumar \latin{et~al.}(2011)Kumar, Alfonso-Prieto, Rovira, Lodowski,
  Jaworska, and Kozlowski]{vitb12-kumar2011}
Kumar,~N.; Alfonso-Prieto,~M.; Rovira,~C.; Lodowski,~P.; Jaworska,~M.;
  Kozlowski,~P.~M. {Role of the axial base in the modulation of the
  cob(I)alamin electronic properties: Insight from QM/MM, DFT, and CASSCF
  calculations}. \emph{J. Chem. Theory Comput.} \textbf{2011}, \emph{7},
  1541--1551\relax
\mciteBstWouldAddEndPuncttrue
\mciteSetBstMidEndSepPunct{\mcitedefaultmidpunct}
{\mcitedefaultendpunct}{\mcitedefaultseppunct}\relax
\EndOfBibitem
\bibitem[Huta \latin{et~al.}(2012)Huta, S., Zgierski, and
  Koz{\l}owski]{vitb12-allouche2012}
Huta,~J.; S.,~P.; Zgierski,~Z.; Koz{\l}owski,~P.~M. {Performance of DFT in
  Modeling Electronic and Structural Properties of Cobalamins}. \emph{J.
  Comput. Chem.} \textbf{2012}, \emph{32}, 174--182\relax
\mciteBstWouldAddEndPuncttrue
\mciteSetBstMidEndSepPunct{\mcitedefaultmidpunct}
{\mcitedefaultendpunct}{\mcitedefaultseppunct}\relax
\EndOfBibitem
\bibitem[Kornobis \latin{et~al.}(2013)Kornobis, Ruud, and
  Kozlowski]{vitb12-kornobis2013}
Kornobis,~K.; Ruud,~K.; Kozlowski,~P.~M. Cob(I)alamin: Insight into the nature
  of electronically excited states elucidated via quantum chemical computations
  and analysis of absorption, CD and MCD data. \emph{J.~Phys.~Chem.~A}
  \textbf{2013}, \emph{117}, 863--876\relax
\mciteBstWouldAddEndPuncttrue
\mciteSetBstMidEndSepPunct{\mcitedefaultmidpunct}
{\mcitedefaultendpunct}{\mcitedefaultseppunct}\relax
\EndOfBibitem
\bibitem[Kumar and Kozlowski(2017)Kumar, and Kozlowski]{vitb12-kumar2017}
Kumar,~M.; Kozlowski,~P.~M. {Electronic and structural properties of
  Cob(I)alamin: Ramifications for B12-dependent processes}. \emph{Coord. Chem.
  Rev.} \textbf{2017}, \emph{333}, 71--81\relax
\mciteBstWouldAddEndPuncttrue
\mciteSetBstMidEndSepPunct{\mcitedefaultmidpunct}
{\mcitedefaultendpunct}{\mcitedefaultseppunct}\relax
\EndOfBibitem
\bibitem[{Dunning Jr.}(1989)]{basis_dunning}
{Dunning Jr.},~T. {Gaussian basis sets for use in correlated molecular
  calculations. {I. T}he atoms boron through neon and hydrogen}. \emph{J. Chem.
  Phys.} \textbf{1989}, \emph{90}, 1007--1023\relax
\mciteBstWouldAddEndPuncttrue
\mciteSetBstMidEndSepPunct{\mcitedefaultmidpunct}
{\mcitedefaultendpunct}{\mcitedefaultseppunct}\relax
\EndOfBibitem
\bibitem[jmo()]{jmol}
Jmol: An Open-Source Java Viewer for Chemical Structures in 3D. {\tt
  http://www.jmol.org/}\relax
\mciteBstWouldAddEndPuncttrue
\mciteSetBstMidEndSepPunct{\mcitedefaultmidpunct}
{\mcitedefaultendpunct}{\mcitedefaultseppunct}\relax
\EndOfBibitem
\bibitem[ein()]{einsum2}
See {\tt https://github.com/jackkamm/einsum2} for more information about the
  \texttt{einsum2} project (accessed October 5, 2020)\relax
\mciteBstWouldAddEndPuncttrue
\mciteSetBstMidEndSepPunct{\mcitedefaultmidpunct}
{\mcitedefaultendpunct}{\mcitedefaultseppunct}\relax
\EndOfBibitem
\bibitem[cup()]{cupy}
See {\tt https://cupy.dev2} for more information about the \texttt{cupy}
  project (accessed October 6, 2020)\relax
\mciteBstWouldAddEndPuncttrue
\mciteSetBstMidEndSepPunct{\mcitedefaultmidpunct}
{\mcitedefaultendpunct}{\mcitedefaultseppunct}\relax
\EndOfBibitem
\bibitem[Gomes and Jacob(2012)Gomes, and Jacob]{gomes_rev_2012}
Gomes,~A. S.~P.; Jacob,~C.~R. Quantum-chemical embedding methods for treating
  local electronic excitations in complex chemical systems. \emph{Annu. Rep.
  Prog. Chem., Sect. C} \textbf{2012}, \emph{108}, 222--277\relax
\mciteBstWouldAddEndPuncttrue
\mciteSetBstMidEndSepPunct{\mcitedefaultmidpunct}
{\mcitedefaultendpunct}{\mcitedefaultseppunct}\relax
\EndOfBibitem
\end{mcitethebibliography}
\end{document}